\documentclass[preprint,12pt,3p]{elsarticle}

\usepackage{matlab-prettifier}
\usepackage{graphicx}
\usepackage{makecell}
\usepackage{xfrac}
\usepackage{parskip}
\usepackage{adjustbox}
\usepackage{setspace}
\usepackage{lscape}
\usepackage[backref=page]{hyperref}
\hypersetup{ 
backref=true, 
bookmarks=true, 
pdftoolbar=true, 
pdfmenubar=true, 
pdffitwindow=false, 
pdfstartview={FitH}, 
pdftitle={}, 
pdfauthor={}, 
pdfsubject={}, 
pdfkeywords={}{}, 
pdfnewwindow=true, 
colorlinks=true, 
linkcolor=BlueViolet, 
citecolor=maroon, 
urlcolor=blue,
}
\usepackage{comment}
\usepackage{subfig}
\usepackage{caption}
\usepackage{longtable}
\usepackage{amssymb}
\usepackage{amsmath}
\usepackage{gensymb}
\usepackage{siunitx}

\usepackage{natbib}
\biboptions{comma,round}
\setcitestyle{authoryear}

\begin{document}
\begin{frontmatter}

\title{The Weitzman premium on the social cost of carbon}

\author[0]{Jinchi Dong}
\author[1,2,3,4,5,6]{Richard S.J. Tol\corref{cor1}
}
\author[7]{Fangzhi Wang}
\address[0]{School of Economics HUST, Huazhong University of Science and Technology, Wuhan, China}
\address[1]{Department of Economics, University of Sussex, Falmer, United Kingdom}
\address[2]{Institute for Environmental Studies, Vrije Universiteit, Amsterdam, The Netherlands}
\address[3]{Department of Spatial Economics, Vrije Universiteit, Amsterdam, The Netherlands}
\address[4]{Tinbergen Institute, Amsterdam, The Netherlands}
\address[5]{CESifo, Munich, Germany}
\address[6]{Payne Institute for Public Policy, Colorado School of Mines, Golden, CO, USA}
\address[7]{Faculty of Finance, City University of Macau, Taipa, Macao}

\cortext[cor1]{Jubilee Building, BN1 9SL, UK}
\fntext[label7]{}

\ead{r.tol@sussex.ac.uk}
\ead[url]{http://www.ae-info.org/ae/Member/Tol\_Richard}

\begin{abstract}
Preference heterogeneity massively increases the social cost of carbon. We call this the Weitzman premium. Uncertainty about an exponential discount rate implies a hyperbolic discount rate, which in the near term is equal to the average discount rate but in the long term falls to the minimum discount rate. We generalise Weitzman's (2001, AER) gamma discounting to zero-inflation and two dimensions but find that the analytical solution is a poor approximation of the non-parametric heterogeneity. We calibrate the pure rate of time preference and the inverse of the elasticity of intertemporal substitution of 79,273 individuals from 76 countries and compute the corresponding social cost of carbon. Compared to the social cost of carbon for average time preferences, the average social cost of carbon is 6 times as large in the base calibration, and up to 200 times as large in sensitivity analyses.
\\
\textit{Keywords}: social cost of carbon, climate policy, discount rate, preference heterogeneity\\
\medskip\textit{JEL codes}: Q54
\end{abstract}

\end{frontmatter}

\newpage \section{Introduction}
The social cost of carbon is a yardstick for the desirable intensity of climate policy. It measures the incremental damage done by emitting a small, additional amount of carbon dioxide. As greenhouse gas emissions change the climate for a prolonged period, the social cost of carbon is a net present value, which depends greatly on the assumed discount rate.

The conventional discount rate is exponential \citep{Samuelson1937, Koopmans1960, Lancaster1963} but people have argued for a hyperbolic discount rate instead \citep{Ainsley1975, Cropper1991, Arrow2014}. In the former, the weight placed on future damages falls at a \emph{constant} rate, in the latter at a \emph{decreasing} rate. See Figure \ref{fig:gamma}.

The late Martin L. Weitzman proposed an elegant and consistent equation for a declining discount rate \citep[][a special case of Theorem 2 in \citet{Chambers2018}]{Weitzman2001}.\footnote{\citet{Millner2020} finds a qualitatively similar effect if discount rates are randomly revised over time.} The initial discount rate is the \emph{average} across a probability distribution or a population. The discount rate's rate of decline depends on the \emph{spread} of uncertainty or opinions about the appropriate discount rate.\footnote{Weitzman uses the standard \emph{deviation}. \citet{Freeman2015ej} argue in favour of using the standard \emph{error} of the mean, if the discount rate is seen as a positive rather than a normative parameter. We here follow Weitzman as our data is about preferences. See \citet{Arrow1996IPCC} and \citet{MillnerHeal2023} for discussions about normative and positive discounting.} Weitzman's formulation is intuitive. People with a high discount rate have essentially no interest in what might happen in the distant future. They place virtually zero weight on events in the future. People with a low discount rate do care about what happens in the long run. Gamma discounting largely disregards the opinion of the indifferent and emphasizes the preferences of those who hold a stake. In the short run, everyone's opinion matters equally; in the long run, the views of the myopic carry less weight. Indeed, in the very long run, a dictatorship of the patient takes hold \citep[][see also Remark 2 in \citet{Chambers2018}]{Milgrom2024}.\footnote{Impatient people are not ignored in the short run, satisfying non-dictatorship as defined by \citet{FengKe2018}.}

Weitzman calibrated the declining discount rate to the preferences of 2,160 PhD economists. Economists may be more financially literate than the general population \citep{Lusardi2014} but their preferences and attitudes are not representative \citep{Carter1991, Dong2024}. We instead use data for the time and risk preferences of almost 80,000 people, consisting of representative samples for 76 countries \citep{Falk2018, Falk2023}. These data are for private decisions in the short run and over small lotteries. We use the Falk data to recalibrate results from \citet{Drupp2018}, which are for public decisions in the long run and for large gambles. The calibration is for the pure rate of time preference $\rho$ and the elasticity of the marginal utility of consumption $\eta$, the two key parameters in the Ramsey rate of discount \citep{Ramsey1928, Arrow2013} $r=\rho+\eta g$, where $r$ is the discount rate and $g$ the growth rate of consumption. Falk's is the largest dataset of its kind, the only one that comes close to claiming global representativeness. Other studies have a smaller number of countries, publish only a few summary statistics per country, or limit themselves to one parameter \citep{Frederick2002, Gandelman2015, Havranek2015JIE, Havranek2015JEEA, Wang2016JEP, Thimme2017, Drupp2018,  Kot2022}.

For reasons explained below, we do not fit a distribution to the preference data but instead use the empirical distribution. That is, we calculate the social cost of carbon for each respondent's $\rho_i$ and $\eta_i$ and contrast this to the social cost of carbon calculated using the average preferences $\bar{\rho}$ and $\bar{\eta}$ for each of the 76 countries as well as the whole world. We call the difference the \emph{Weitzman premium}: $W = \overline{SCC(\rho, \eta)} - SCC(\bar{\rho},\bar{\eta})$. The Weitzman premium is large.

We use a globally aggregated model. This allows us to focus on the Weitzman premium, which reflects heterogeneity \emph{in preferences}. There is, of course, heterogeneity in income, vulnerability, and so on. We could have included that in the model\textemdash see \citet{Fankhauser1997} and \citet{Anthoff2010} for a discussion of the aggregation of impact over people with different incomes and \citet{Tol2019EE} and \citet{Tol2021anyas} for a treatise on heterogeneity in vulnerability to climate change\textemdash but then we would have had to disentangle these three premia on the social cost of carbon to isolate the effect of preference heterogeneity.

The next three sections present theory and methods and discuss results. The final section concludes. Additional material is in the appendix.

\section{Theory}
\label{sc:theory}
Martin L. Weitzman considered discounting under uncertainty about the discount rate. \citet{Weitzman1998discount} showed that the discount rate converges over time to its lower bound. \citet{Weitzman2001} showed that, if the uncertainty about the discount rate follows a Gamma distribution, then the discount rate is a function of time $r(t) = \frac{\alpha}{\beta + t}$, where $alpha$ is the shape parameter and $\beta$ is the rate parameter. As $\mu = \frac{\alpha}{\beta}$, the effective discount rate is its mean at the start $t=0$ and falls to zero, the lower bound of the Gamma distribution, in the long run $t \rightarrow  \infty$. \citet{Weitzman2001} applies this to the consumption discount rate and a panel of 2160 economists who expressed an opinion on the appropriate social discount rate. Although Weitzman interprets this as \emph{uncertainty} about the discount rate but it can it also be interpreted as heterogeneity in taste.

We extend Weitzman's analysis in a number of ways but stay true to the spirit of his proposal. The difference between the variable of interest, here the social cost of carbon, with a constant and a declining discount rate is dubbed the Weitzman premium. It is a premium that originates from heterogeneity in taste.

First, we consider taste-heterogeneity in the general population rather than among economists. This does not require a methodological extension. Rows 1 and 4 of Table \ref{tab:analytics} reveals that a representative sample of the world population has a higher discount rate, and so a lower social cost of carbon, than Weitzman's convenience sample of economists.

Rows 2 and 5 of Table \ref{tab:analytics} confirm that Gamma discounting raises the social cost of carbon. This is as expected. The Gamma discount rate starts at the sample mean and falls to zero over time. More weight is thus placed on the future. The Weitzman or taste-heterogeneity premium is the relative increase in the social cost of carbon due to a declining discount rate. It is 70\% for Weitman's data, 117\% for Falk's data.

Second, we note that some people advocate a discount rate of zero. The Gamma distribution has no probability mass at zero. A zero-inflated Gamma distribution implies a simple extension:
\begin{equation}
    r(t) = (1-\delta) \frac{\alpha}{\beta + t}
\end{equation}
where $1-\delta$ is the fraction of respondents who have a positive discount rate.

Rows 5 and 6 of Table \ref{tab:analytics} shows that this \emph{decreases} the social cost of carbon. The reason is that, although 5\% of the sample does not discount the future at all, the remaining 95\% discount the future harder at the start (7.1\% v 7.5\%) and the discount rate declines at a slower rate ($\beta = 34$ v $\beta = 39$).

Third, we use the Ramsey rule $r = \rho + \eta g$. This varies the discount rate with economic growth. Rows 4 and 7 of Table \ref{tab:analytics} show that the social cost of carbon is lower for Ramsey discount rate. This is because the constant discount rate in row 4 is calibrated to the sluggish growth of the world economy in the recent past, whereas the Ramsey discount rate in row 7 uses the optimistic growth rate in the SSP2 scenario. Faster growth implies a higher discount rate and a lower social cost of carbon.

Fourth, if $\rho$ and $\eta$ are independent and Gamma distributed
\begin{equation}
    r(t) = \frac{\alpha_\rho}{\beta_\rho + t} + \frac{\alpha_\eta g}{\beta_\eta + g t}
\end{equation}
We derive this result in the Appendix and show that this discounts future payoffs more than Weitzman's Gamma discounting for a consistent calibration. Row 8 in Table \ref{tab:analytics} shows that the Weitzman premium indeed increases, to 143\%.

Fifth, there is no reason to assume that $\rho$ and $\eta$ are independent. In that case,
\begin{equation}
    r(t) = \frac{\alpha_\rho (1+\kappa g)}{\beta_\rho + t(1 + \kappa g)} - \sigma^2 g t
\end{equation}
where $\kappa$ is the estimate coefficient of a regression of $\eta$ on $\rho$ and $\sigma$ is the standard error that regression. This equation has a problem. As noted by \citet{Weitzman1998discount}, the first three elements tend to the minimum in the sample\textemdash just above zero in this case. The fourth element, $-\sigma^2 g t$ tends to minus infinity. In the long run, the discount rate turns negative\textemdash in our base scenario, this happens well before 2100. The net present value is thus unbounded.

We note just how complicated things get, and that computers are much faster than when Weitzman derived his theorem. We can consider heterogeneity in taste by assuming a distribution to describe the data in our sample and derive the behaviour of the discount rate under that distributional assumption. We can also just compute the discount rate for everyone in our sample at every point in the future and take the average. That is, sixth, we use the empirical distribution.

Comparing rows 2 and 3 in Table \ref{tab:analytics} shows that Weitzman's Gamma distribution closely approximates the empirical distribution. The difference is roughly 5\%. However, rows 8 and 10 show that this does not carry over to the Ramsey rule. The analytical approximation (without zero-inflation, without correlation) has a social cost of carbon of \$13.2/tC. The social cost of carbon using the empirical distribution is \$30.5/tC. There is a positive correlation between the pure rate of time preference and the inverse of the elasticity of intertemporal substitution. This shifts probability mass towards low and high discount rates; see Figure \ref{fig:correl}. But, as Weitzman showed, high discount rates are largely ignored in the long run. The result is a considerable Weitzman premium, over 450\%.

\section{Application}
\emph{Model}\textemdash We use a standard integrated assessment model, combining models of the carbon cycle \citep{MaierReimer1987}, climate \citep{Schneider1981}, economic growth \citep{Solow1956}, and impacts \citep{Barrage2024}.

\emph{Impacts}\textemdash We use the latest version of \textsc{dice} to model the impacts of climate change \citep{Barrage2024}. As a sensitivity analysis, we use an earlier meta-analytic impact function \citep{Howard2017} and four alternative functions from a more recent meta-analysis \citep{Tol2024EnPol}. In contrast to most integrated assessment models, impacts of climate change fall as per capita income increases \citep{Schelling1992}, using an income elasticity of -0.36 \citep{Botzen2021}. As a sensitivity analysis, we half and double this, set it to zero, and try a small, positive income elasticity. Recall that we model the global aggregate, so the income elasticity applies over time only.

\emph{Scenarios}\textemdash We use SSP2 as our base case as this is most likely scenario \citep{Srikrishnan2022}. The other SSPs and the earlier SRES scenarios are used as sensitivity analyses.

\emph{Preferences}\textemdash We use stated preferences from an unincentivized survey. The unincentivized survey measures were validated with an incentivized survey and an incentivized experiment \citep{Falk2018, Falk2023}. 80,337 replied to the survey \citep{Falk2018, Falk2023}. Removing missing observations, 79,273 respondents remain.

A generic question about time\textemdash how willing are you to give up something that is beneficial for you today in order to benefit more from that in the future?\textemdash is combined with a decision tree with five layers about trade-offs between today and a year from now. Answers are combined in an index between -1.3 and +2.8.

The decision tree is about the consumption discount rate over a short period in which income growth can reasonably be expected to be small. The consumption discount rate is thus close to the utility discount rate or pure rate of time preference. The expressed time preference is private. Private time preferences differ from social ones, but preciously little is known about how much \citep{Percoco2006, Mahboub-Ahari2014}. It stands to reason that privately impatient people are also impatient socially. We therefore use the data by \citet{Drupp2018} on social time preference in the long run to calibrate Falk's time index.

Similarly, Falk's risk index, ranging from -1.9 to +2.5, is based on a generic question\textemdash generally speaking, are you a person who is willing to take risks or do you try to avoid risks?\textemdash and a five-layered decision tree between risky and less risky lotteries.

As above, we use Drupp's data to calibrate Falk's private preferences to social ones. The utility function is differently curved in risk and time space \citep{Epstein1991}. We assume that the parameter that governs trade-offs between a fortunate and unfortunate self is correlated to the parameter for trade-offs between a richer future self and a poorer current one. Again, we do not use Falk's risk index directly. We use it for calibration only.

The national samples are not representative of country populations but sampling weights are available to make the sample representative on observed characteristics. Unfortunately, only part of the data is publicly available. Demographic information only includes age and gender. Although there are six questions each on time and risk, only the composite time and risk indices have been released.

\emph{Calibration}\textemdash We discount marginal impacts using the Ramsey rule \citep{Ramsey1928}, which states that the discount rate $r$ equals the pure rate of time preference $\rho$ plus the growth rate of consumption $g$ times the elasticity of marginal utility of consumption $\eta$: $r = \rho + \eta g$. The survey of about 80,000 laypeople has \emph{indices} of time and risk preferences. However, another survey has data on the parameters of the Ramsey rule \citep{Drupp2018}. This is a survey of some 200 professional economists, who understand the application of the Ramsey rule to public policy. We linearly calibrate the national average time and risk average to the economists' survey so that the fifth and ninety-fifth percentiles match. Ninety percent of the imputed preferences are therefore within the observed range. We further ensure that $\rho \geq 0$ and $\eta \geq 0$.

\citet{Dong2024} test the robustness of this calibration. We use the same procedure but, first, we weigh the country averages by the population of the country. In the population-weighted calibration, people are more impatient and more risk-averse, discounting the future harder, and the spread of opinions is larger. In the geographically-restricted calibration, people are less impatient and less risk-average, and the spread of opinions on time discounting is smaller.

Second, we limit the sample of laypeople to countries in Western Europe and North America to match the sample of experts, who are based there. Figure \ref{fig:calib} shows the cumulative density function of the social cost of carbon for the Drupp sample and for the calibrated Falk sample, limited to the countries of origin of Drupp's respondents. There is a good match between the distributions.

\emph{Computation}\textemdash We run the integrated assessment model with the time and risk preference for each individual in the sample; for the weighted average for each country; and for the weighted average for the world. We compare the individual results to an extension of the analytical approach proposed by Weitzman. The qualitative results are the same, but the numerical results are different.

\emph{Validation}\textemdash We use carbon price data from the \textit{Carbon Pricing Dashboard} of the World Bank, data from Climate Action Tracker on emission reduction commitments and emission projections under current policies, and data from recent surveys on support for climate action \citep{Andre2024, dechez2025} to test our estimates of the national cost of carbon.

\section{Results}
Table \ref{tab:global} shows the social cost of carbon for average preference and the average social cost of carbon. The Weitzman premium is the difference between the two. The social cost of carbon is low compared to the literature \citep{Tol2025anyas} because we assume that vulnerability to climate change falls as people grow richer \citep{Diaz2017} and because the average person is more impatient than economists from the Global North \citep{Dong2024}.

\begin{figure}
  \begin{centering}
  \caption{The Weitzman premium on the social cost of carbon, by country}
  \label{fig:premium}
  \includegraphics[width=\textwidth]{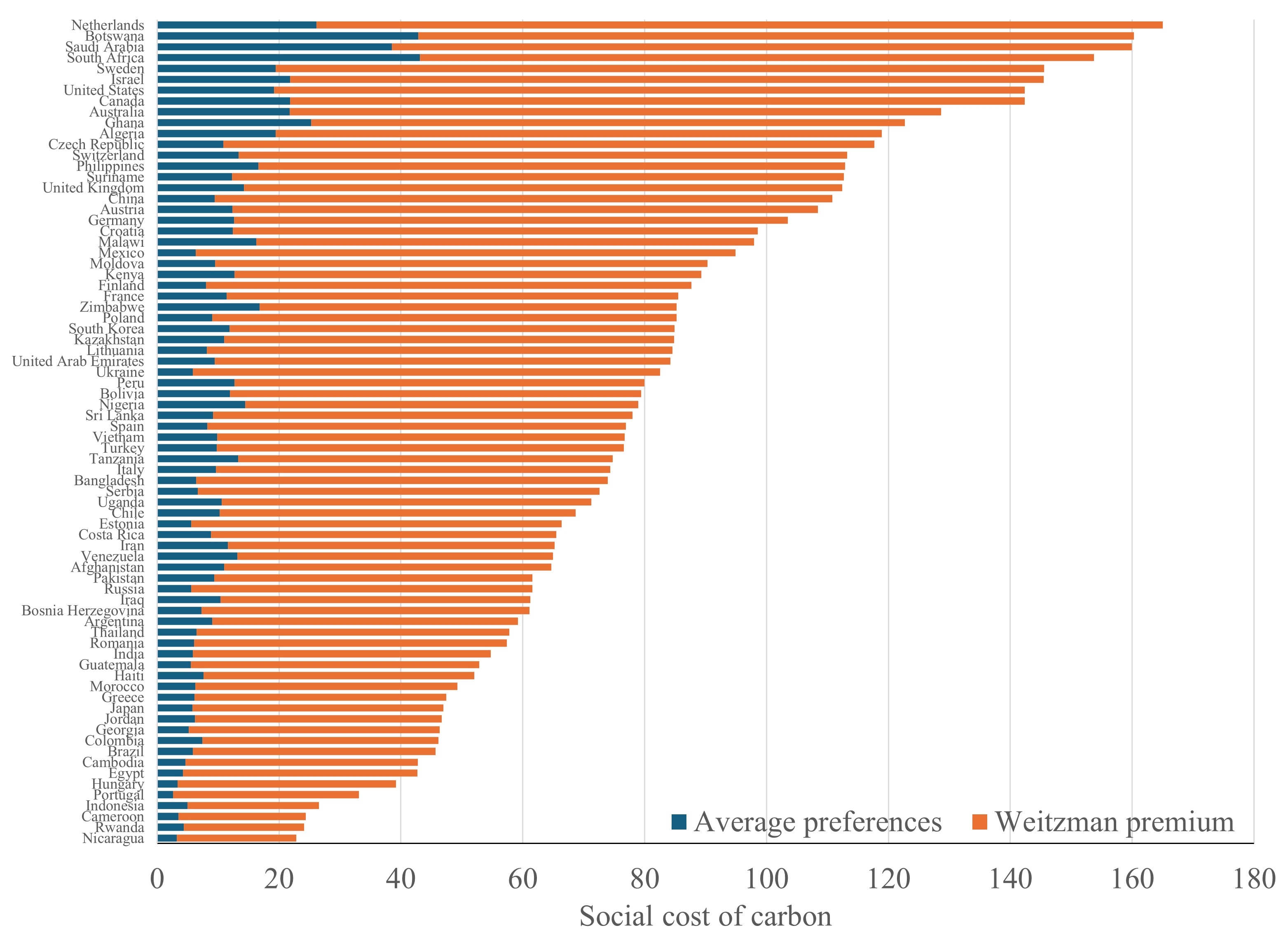}
  \end{centering}
  {\footnotesize The blue bar shows the social cost of carbon for the national average time and risk preferences for the 76 countries in our sample. The blue plus orange bar shows the national average social cost of carbon for individual time and risk preferences. The difference between the two, the orange bar, is the social cost of carbon. All numbers are in 2010 \$ per metric tonne of carbon for carbon dioxide emissions in 2020.}
\end{figure}

The Weitzman premium is large. Figure \ref{fig:premium} shows the social cost of carbon, averaged over the residents of each of the 76 countries using their individual preferences. This is contrasted with the social cost of carbon using the national average preferences. The absolute Weitzman premium for 2020 is smallest in Nicaragua, \$20/tC, and largest in the Netherlands, \$139/tC. Its relative contribution to the social cost of carbon is smallest in South Africa, 72\%, and largest in Mexico, 93\%. That is, the Weitzman premium increases the social cost of carbon by a factor of 3.6 to 15.

At the global average, the Weitzman premium is \$25/tC, making up 82\% of the social cost of carbon, increasing it by a factor of 5.6.

\begin{figure}
  \begin{center}
  \caption{The impact of preferences on the social cost of carbon and the Weitzman premium}
  \label{fig:mechanisms}
  \includegraphics[width=0.49\textwidth]{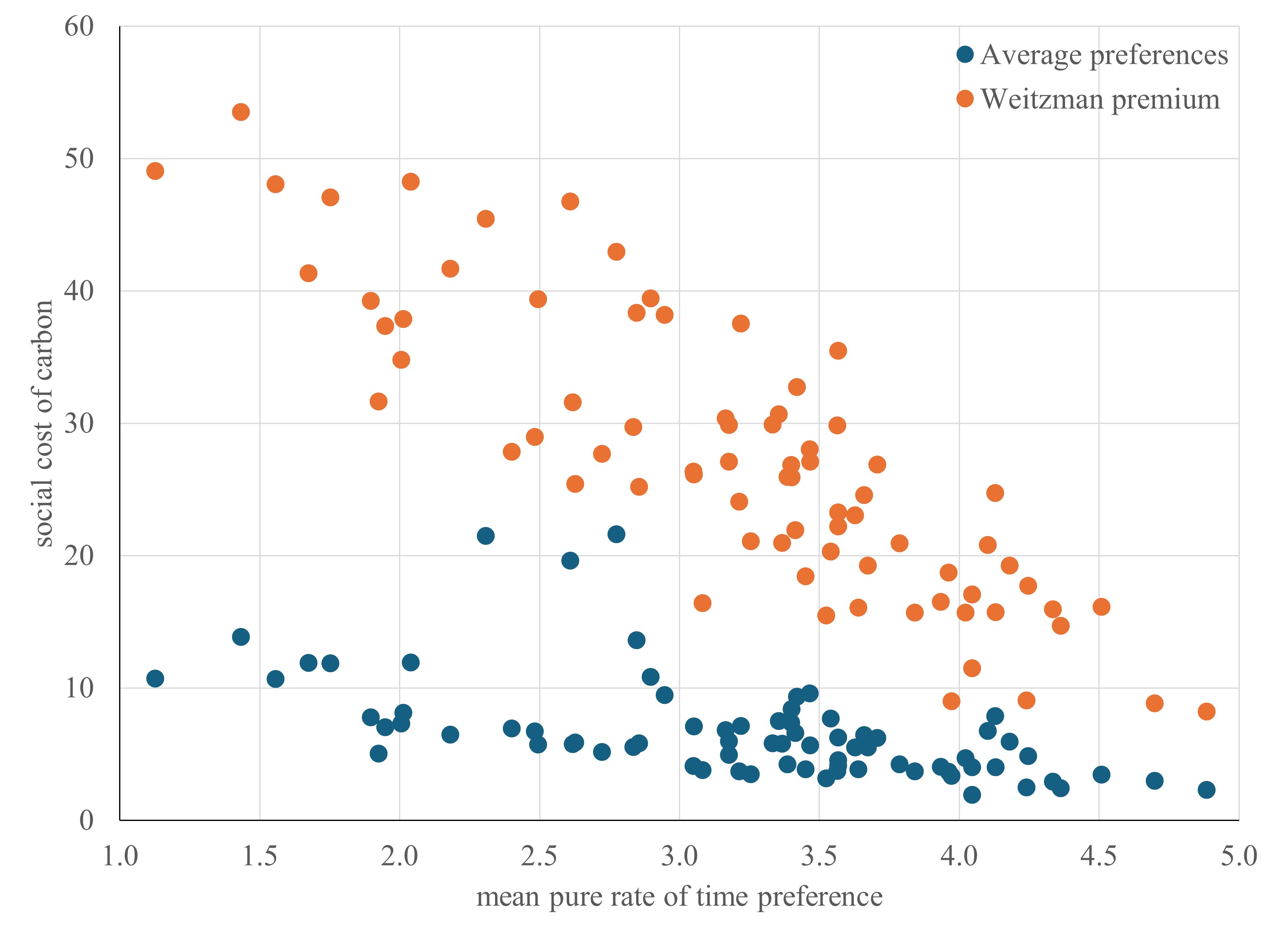}
  \includegraphics[width=0.49\textwidth]{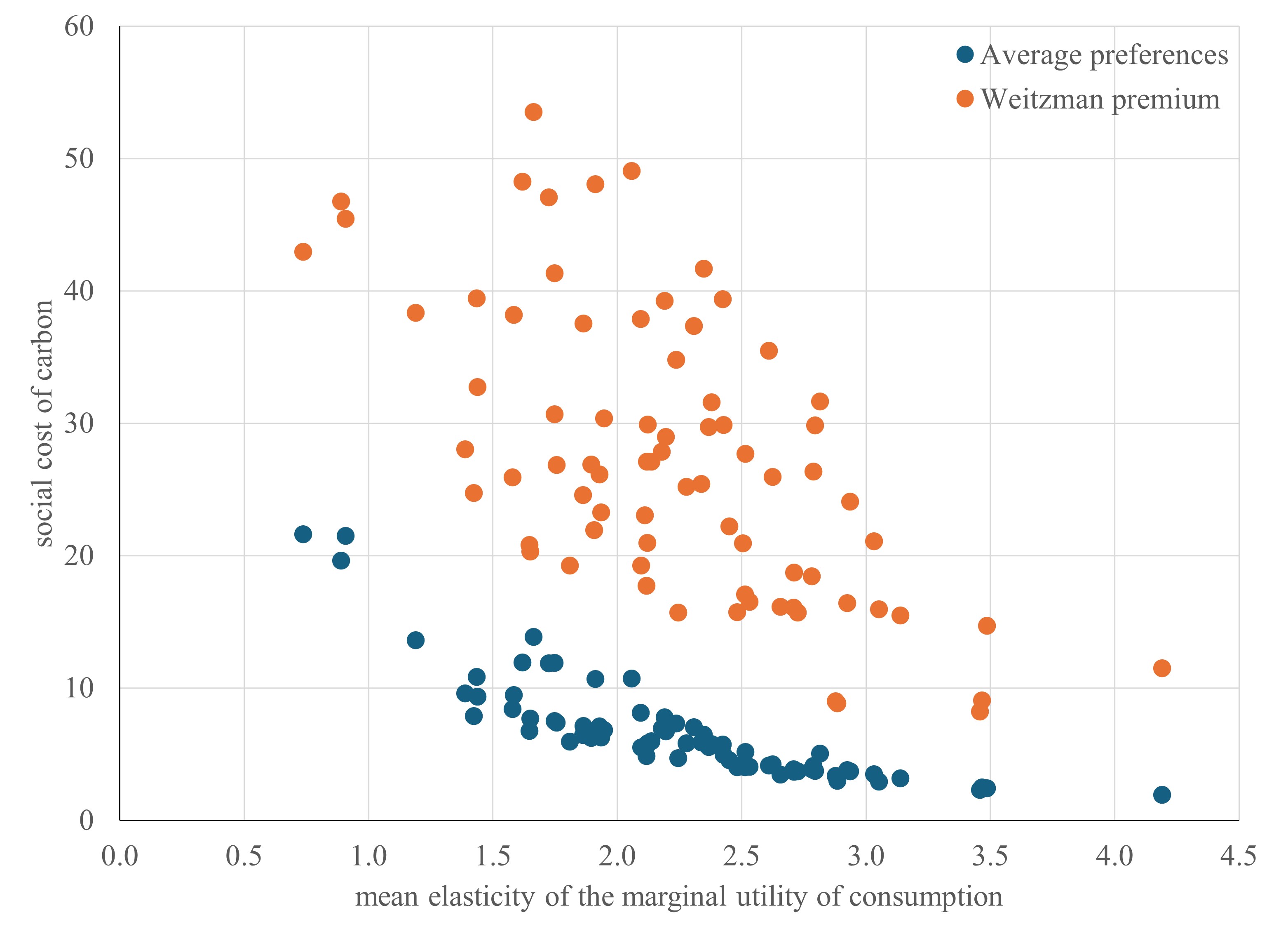}
  \includegraphics[width=0.49\textwidth]{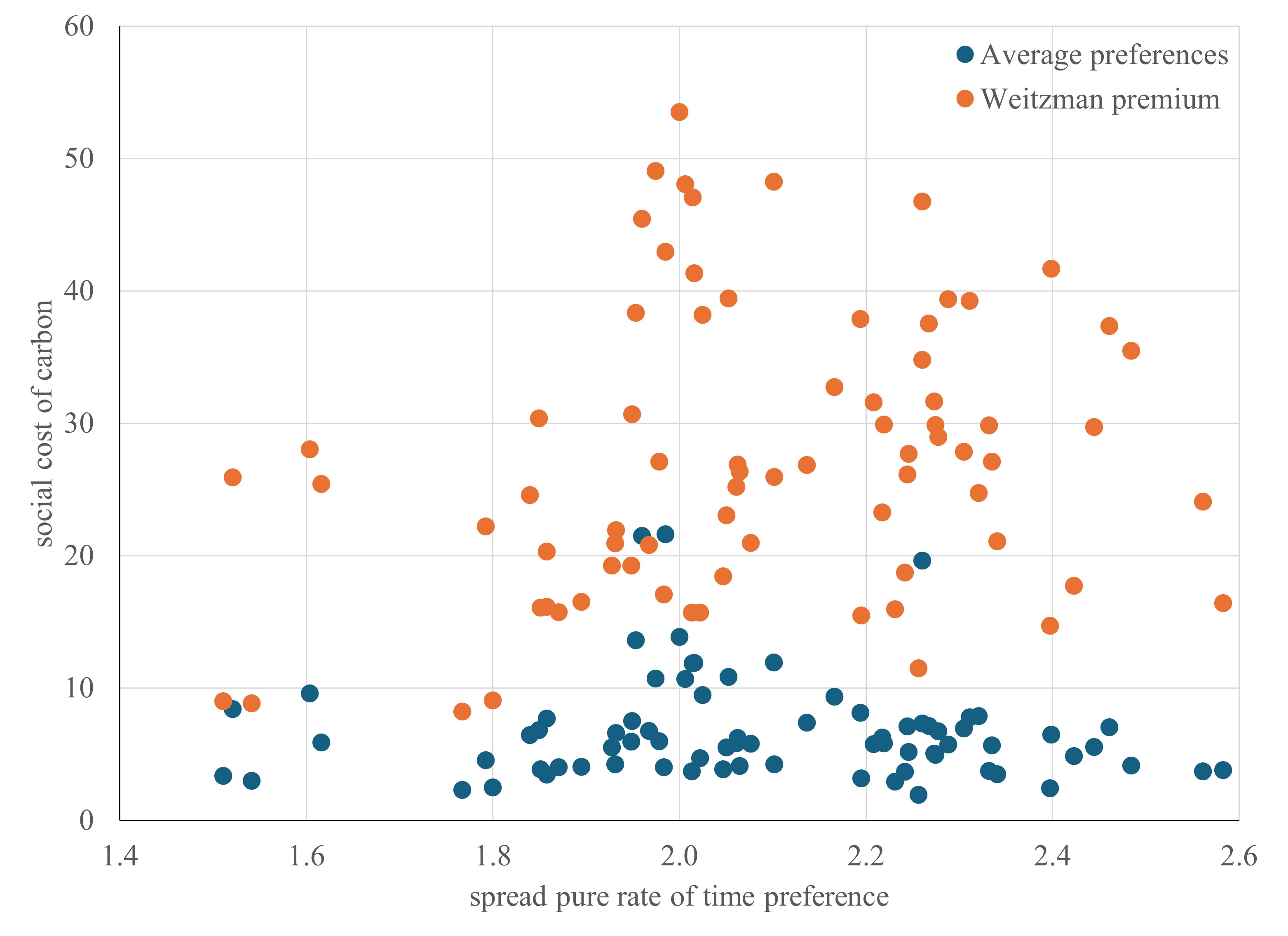}
  \includegraphics[width=0.49\textwidth]{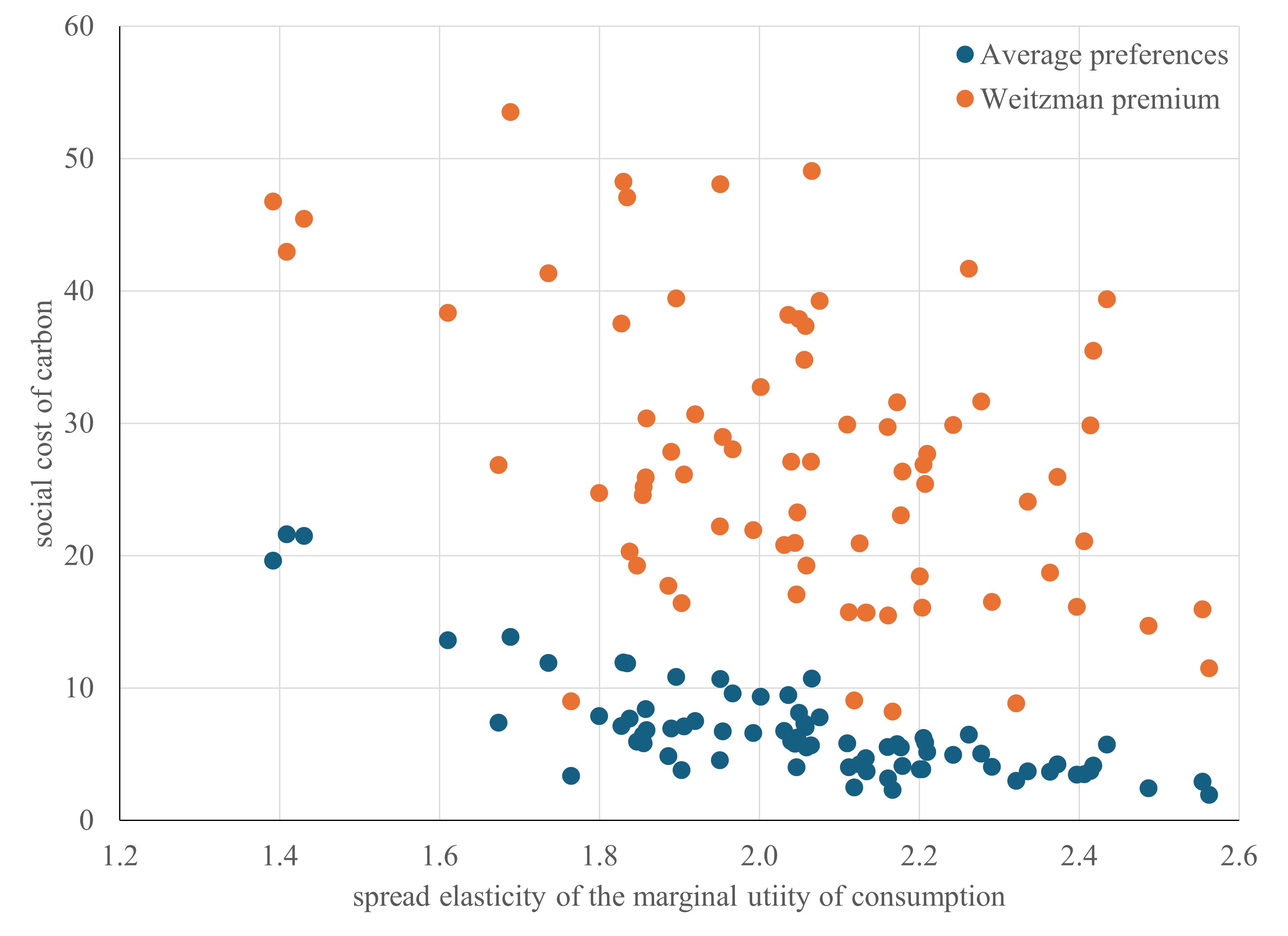}
  \includegraphics[width=0.49\textwidth]{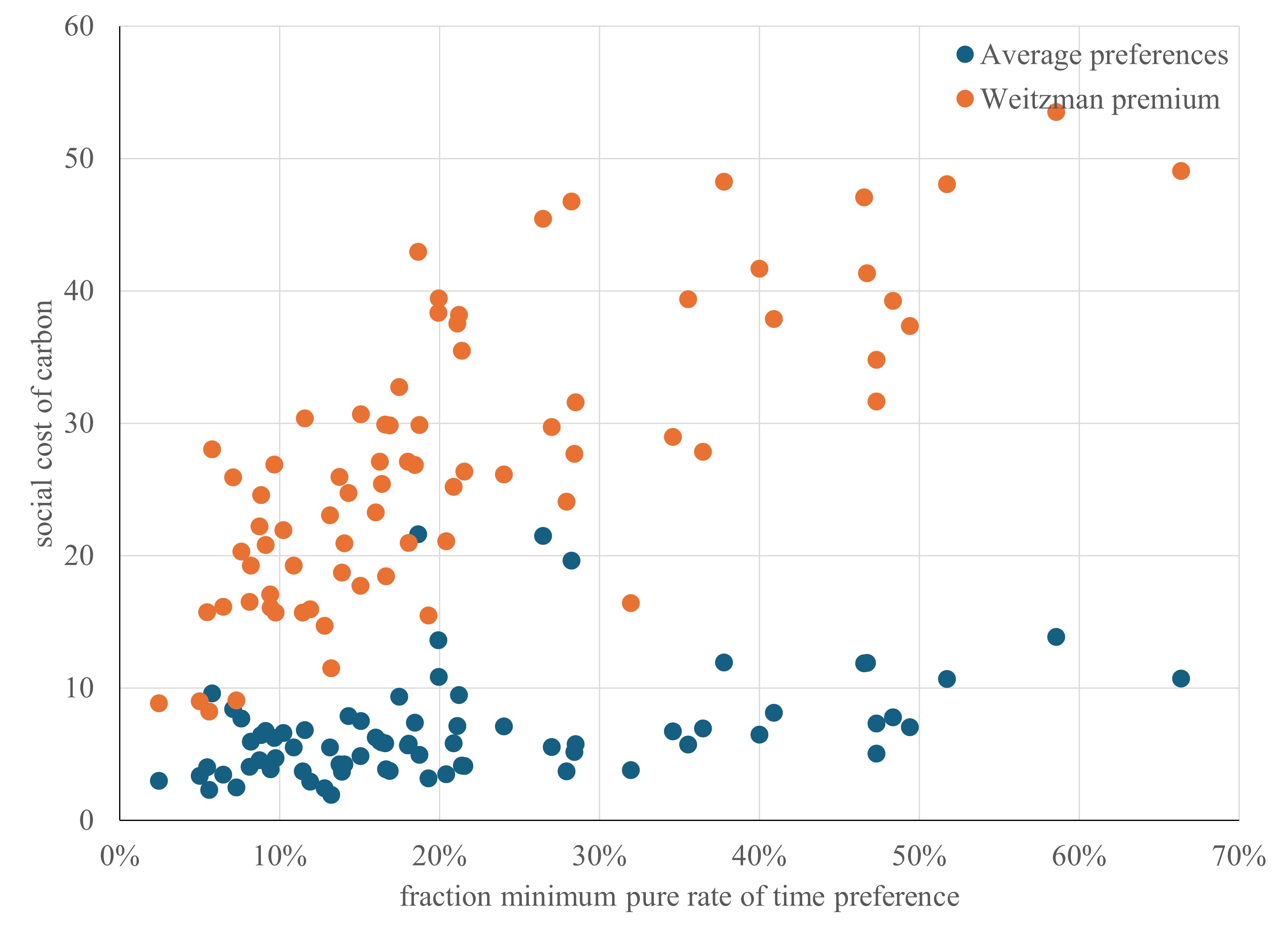}
  \includegraphics[width=0.49\textwidth]{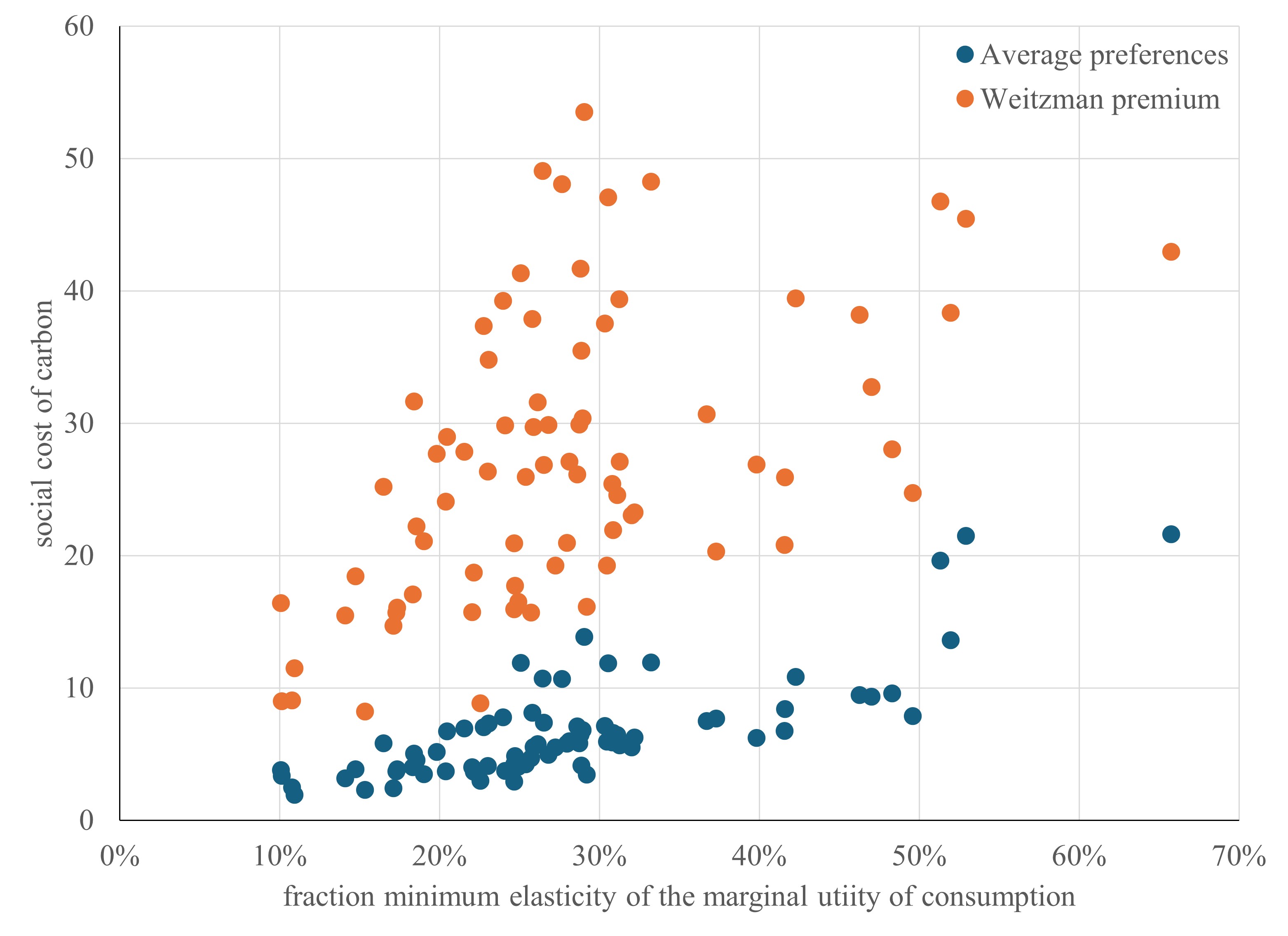}
  \end{center}
  {\footnotesize The blue dots show the social cost of carbon for the national average pure rate of time preference and the national average elasticity of the marginal utility of consumption for the 76 countries in the sample. The orange dots show the corresponding Weitzman premium. These statistics are plotted against the pure rate of time preferences (left column) and the elasticity of the marginal utility of consumption (right column), against the mean (top row), the standard deviation (middle row), and the fraction of zeroes (bottom row).}
\end{figure}

Figure \ref{fig:mechanisms} reveals the mechanisms. The social cost of carbon and the Weitzman premium increase as the average pure rate of time preference in a country's population falls, and as the average elasticity of marginal utility of consumption falls. This is as expected. The Ramsey discount rate falls with these two parameters. A lower discount rate implies greater care about future problems such as climate change.

\begin{figure}
  \begin{center}
  \caption{Time and risk preferences: Mean vs spread}
  \label{fig:meanvspread}
  \includegraphics[width=0.49\textwidth]{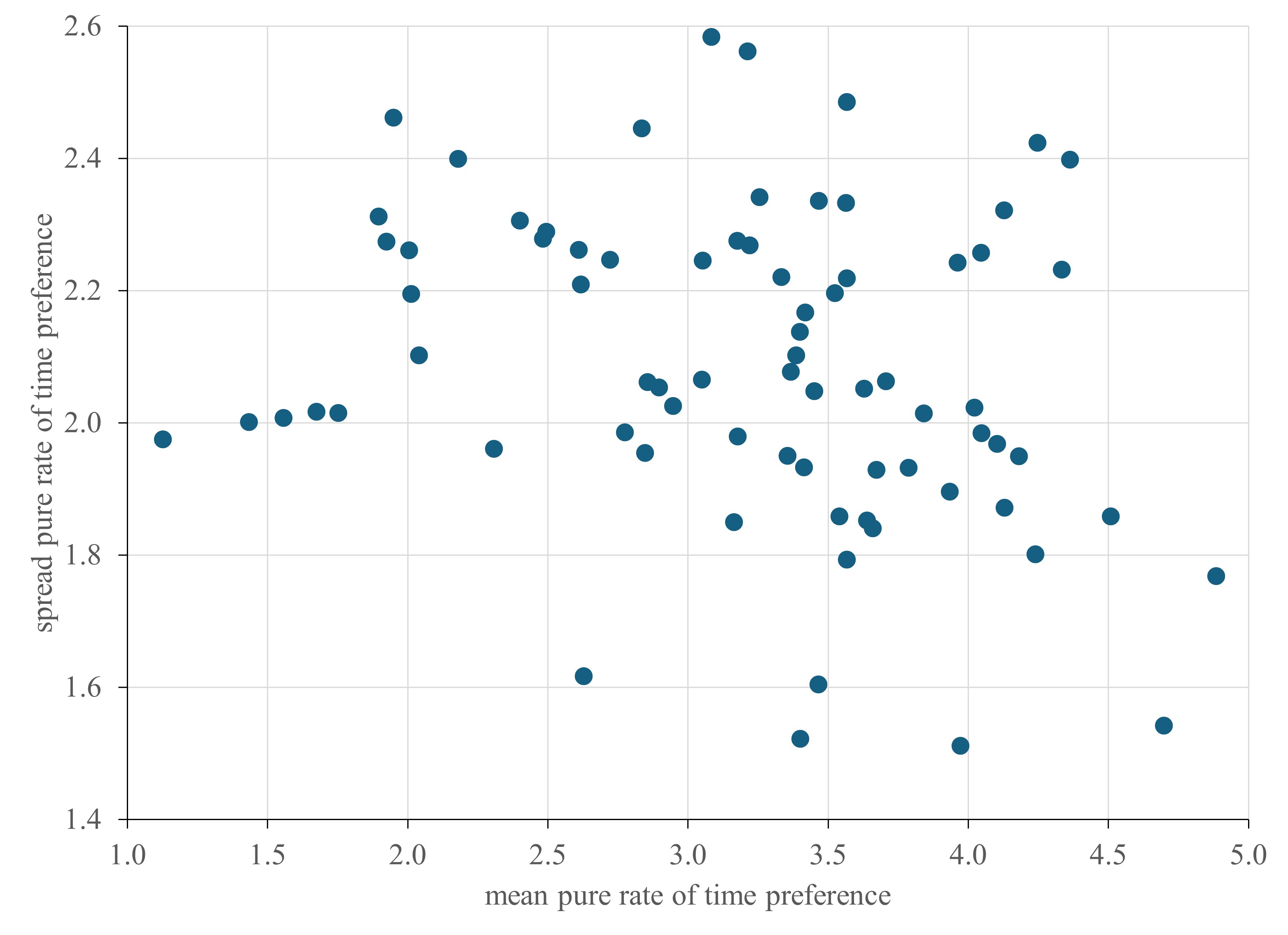}
  \includegraphics[width=0.49\textwidth]{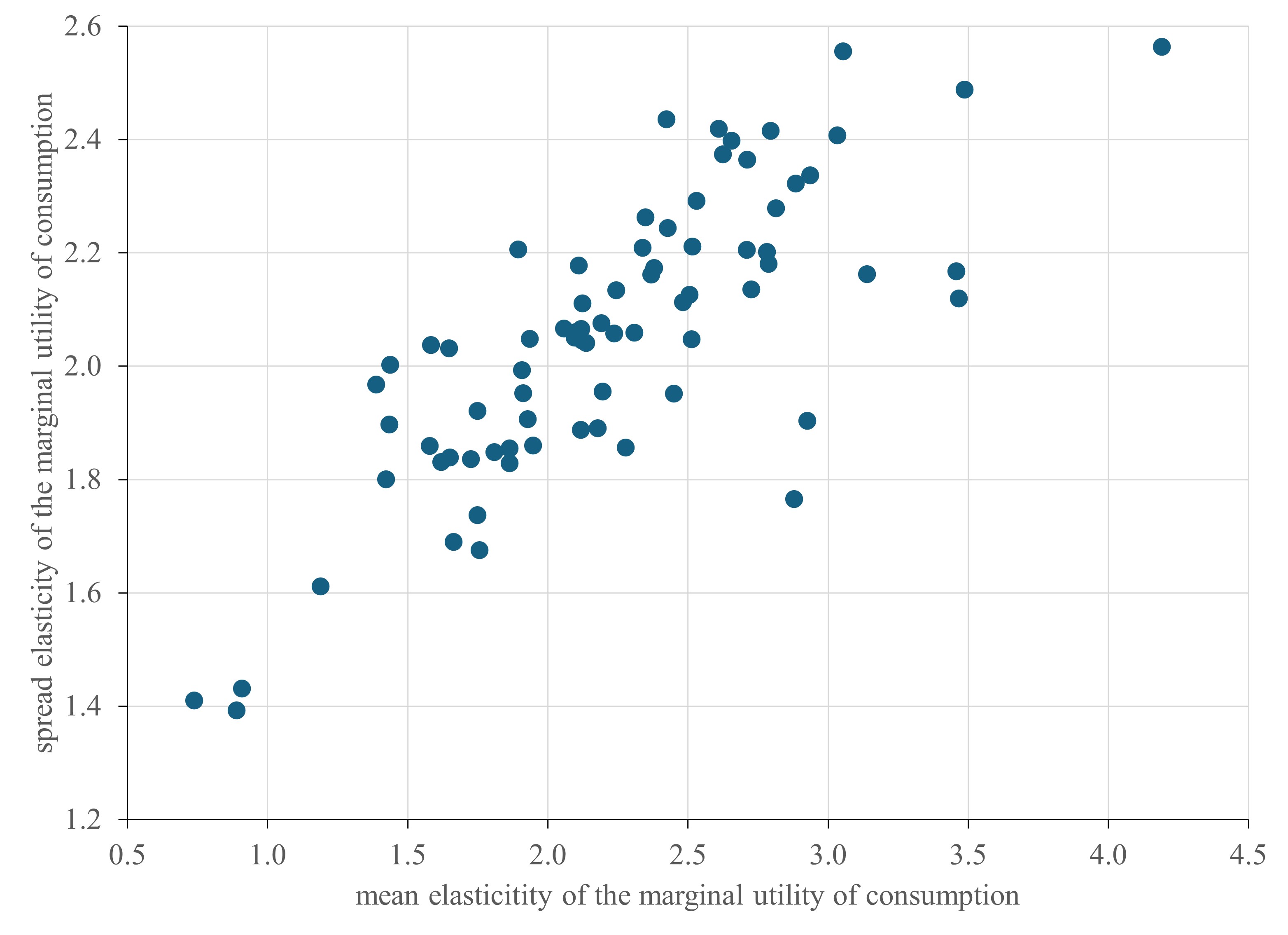}
  \end{center}
  {\footnotesize The left panel plots the average pure rate of time preference per country against the standard deviation of that country. The right panel plots the average elasticity of the marginal utility with respect to consumption against its standard deviation.}
\end{figure}

The spread of time preference does not systematically affect the social cost of carbon. More uniform views on the curvature of the utility function do somewhat raise the social cost of carbon and the Weitzman premium, but this is because the mean falls with the spread (see Figure \ref{fig:meanvspread}). However, the fraction of people who are not myopic, $\rho=0$, or whose utility is linear in consumption, $\eta=0$, increases the social cost of carbon and the Weitzman premium. A general convergence of preferences would not greatly affect the Weitzman premium; convergence of those in the left tail to the mean would. This is as Weitzman predicted \citep{Weitzman1998discount}: What matters, for long-term problems, is the opinions of only the most patient people.

\begin{figure}
  \begin{centering}
  \caption{The social cost of carbon}
  \label{fig:histscc}
  \includegraphics[width=\textwidth]{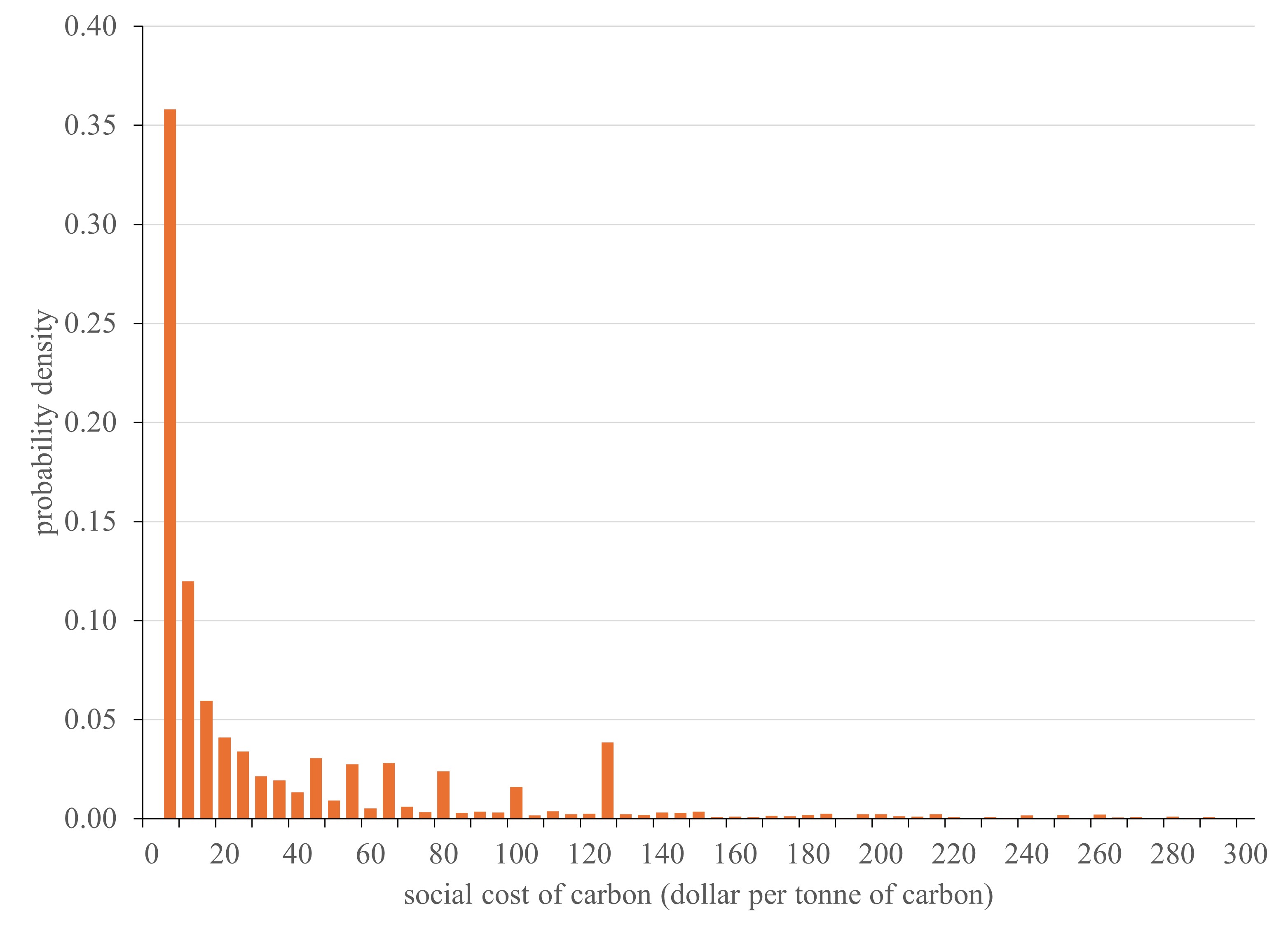}
  \end{centering}
  {\footnotesize Histogram of the social cost of carbon, in 2010 US dollar per metric tonne of carbon, for the individual time and risk preferences of the 79,272 respondents. Results are weighted to be representative of the world population.}
\end{figure}

Figure \ref{fig:histscc} shows the histogram of the social cost of carbon for each of the roughly 80,000 respondents. The distribution of the social cost of carbon is heavily skewed. The mode is \$1/tC, the median \$7/tC, the mean \$31/tC. The mean is at the 77\%ile.

Figure \ref{fig:histscc} counts every individual equally. This reflects direct democracy at a global level\textemdash one person, one vote. Table \ref{tab:global} adds alternative aggregates. The United Nations system is based on one country, one vote, whereas the market is a plutocracy, based on one dollar, one vote. Equity weights correct for the fact that a dollar to a rich woman is not the same as a dollar to a poor woman \citep{Fankhauser1997}. These alternatives are based on the country-average social cost of carbon; there are many ways to add opinions within a jurisdiction, but these are not explored here. All three alternative aggregations raise the Weitzman premium and the social cost of carbon, the UN system to \$34/tC because it favours smaller countries such as Israel and the Czech Republic, the market system to \$43/tC because it favours rich countries such as the Netherlands and Sweden, and the welfare system to \$55/tC because it favours poor countries such as Botswana and Ghana\textemdash the people of these countries all prefer an above average social cost of carbon. The weighted average social cost of carbon tends to be larger than the unweighted average because the social cost of carbon is bounded from below (at zero, by construction) but not bounded from above. Therefore, there are subsamples with really high estimates, but no subsamples with really low estimates to offset the high ones.

Figure \ref{fig:meanvspread} plots, for every country, the average calibrated pure rate of time preference against the standard deviation of the same (left panel). The right panel plots the average calibrated elasticity of marginal utility of consumption against its standard deviation. The left panel reveals that there is no systematic relationship between the mean and spread of time preferences. However, the right panel shows that a sharper curvature of the utility function is associated with a wider variety of opinions about that curvature.

\begin{figure}
  \begin{centering}
  \caption{Sensitivity analysis}
  \label{fig:sensana}
  \includegraphics[width=\textwidth]{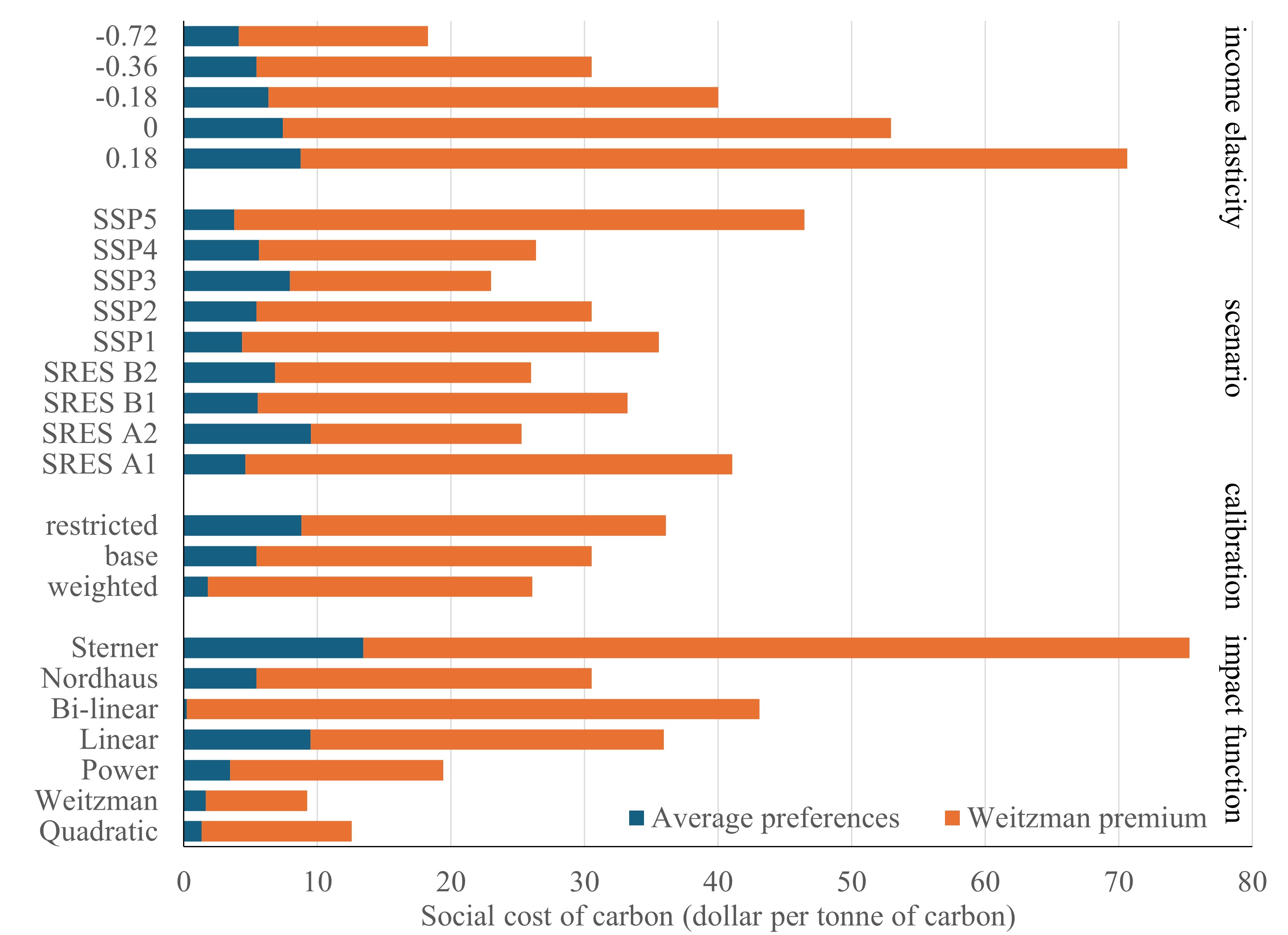}
  \end{centering}
  {\footnotesize Global average social cost of carbon and Weitzman premium for alternative income elasticities of the impact of climate change, alternative scenarios, alternative calibrations, and alternative impact functions.}
\end{figure}

The social cost of carbon measures, at the margin, the severity of \emph{future} impacts and is thus highly uncertain. Figure \ref{fig:sensana} shows a sensitivity analysis around four key uncertainties. The social cost of carbon varies with the choice of impact function, with the method of calibrating preferences, with the choice of scenario, and with the income elasticity; our base assumptions lie somewhere in the middle. The social cost of carbon is lower if vulnerability to climate change falls with economic growth. In every case, the Weitzman premium is substantial. It is very substantial indeed\textemdash a factor 195\textemdash in case of a bi-linear impact function \citep{Tol2009JEP}, which has climate change benefits in the short run (which are emphasized by the high average discount rate) but damages in the long run (which are emphasized by the people with a low discount rate).

In the base calibration, the weighted global average of the social cost of carbon is \$30.5/tC while the social cost of carbon for average preferences equals \$5.4/tC. In the population-weighted (geographically-restricted) calibration, both numbers fall (rise) to \$26.1/tC (\$36.1/tC) and \$1.8/tC (\$8.8/tC), respectively. The absolute Weitzman premium falls (rises) too, from \$25.1/tC to \$24.3/tC (\$27.3/tC), but the relative premium increases from 5.6 to 14.4 (4.1). See Figure \ref{fig:sensana}.

Individual data allow for more than just calculating the Weitzman premium for \emph{countries}. Unfortunately, the public data file only has country of residence, gender, and age. Let us consider the latter two.

The pure rate of time preference is almost the same for men (3.1) and women (3.3). However, women's elasticity of marginal utility with respect to consumption is higher, 2.5 versus 2.0. Men are less risk averse, at least according to the Falk data. Therefore, women have a lower social cost of carbon than men, $\$30(\pm 0.4)$/tC versus $\$39(\pm 0.3)$/tC, a statistically significant difference.

\begin{figure}
  \begin{centering}
  \caption{The social cost of carbon by age}
  \label{fig:age}
  \includegraphics[width=\textwidth]{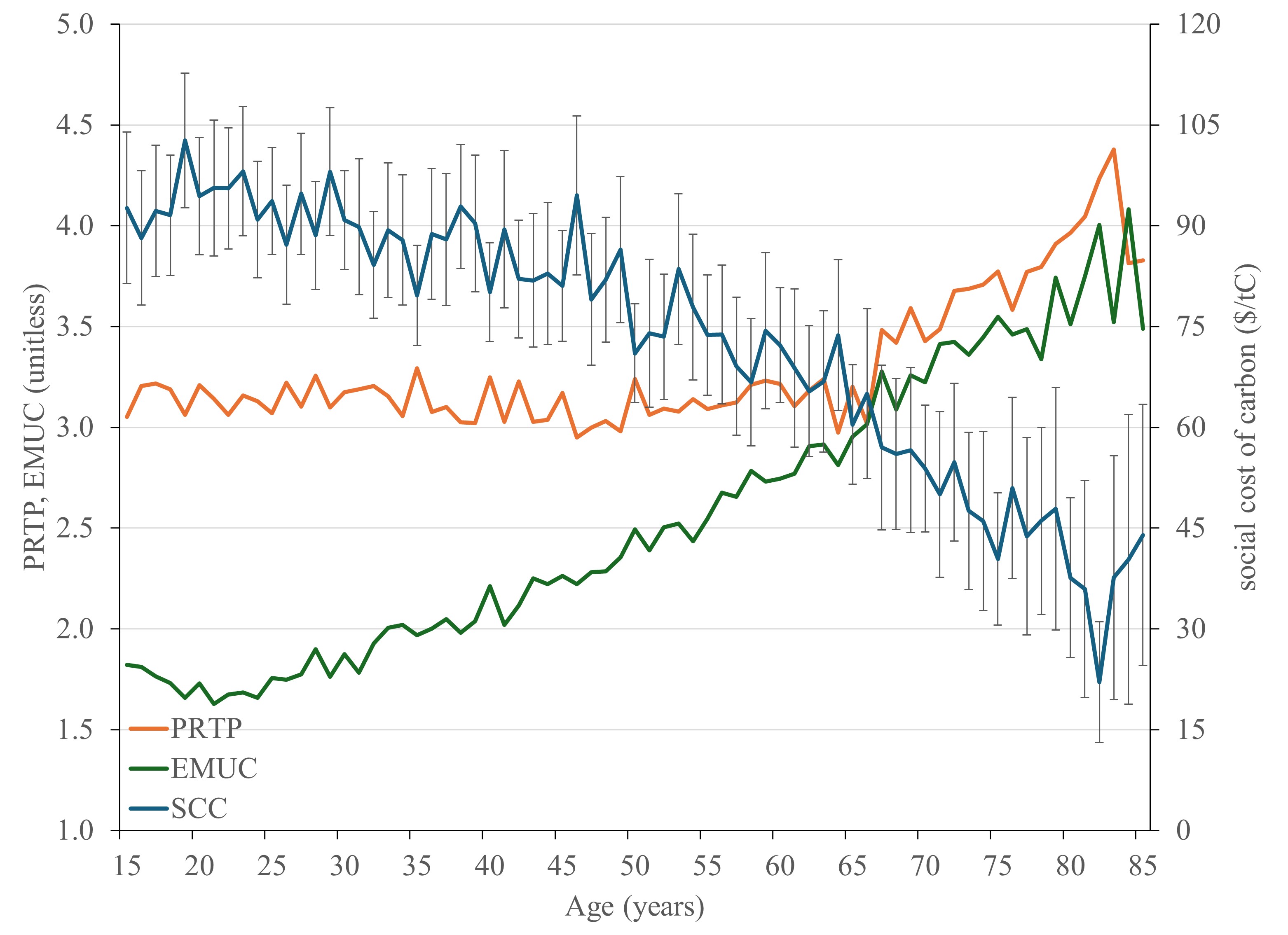}
  \end{centering}
  {\footnotesize The blue line shows the average social cost of carbon by age. The bars denotes the 95\% confidence interval. The orange and green lines show the average pure rate of time preference (PRTP) and elasticity of the marginal utility of consumption (EMUC), respectively, by age.}
\end{figure}

Figure \ref{fig:age} shows the social cost of carbon by age. The pure rate of time preference is the same between ages 15 and 65, but increases for older people. The curvature of the utility function shows a steady increase with age. Therefore, the social cost of carbon is significantly higher for younger people. This is not inconsistent with survey evidence that younger people tend to be more concerned about climate change \citep{Lee2015, Poortinga2019}.

\begin{figure}
  \begin{center}
  \caption{The social cost of carbon plotted against measures of climate policy intensity}
  \label{fig:validation}
  \includegraphics[width=0.49\textwidth]{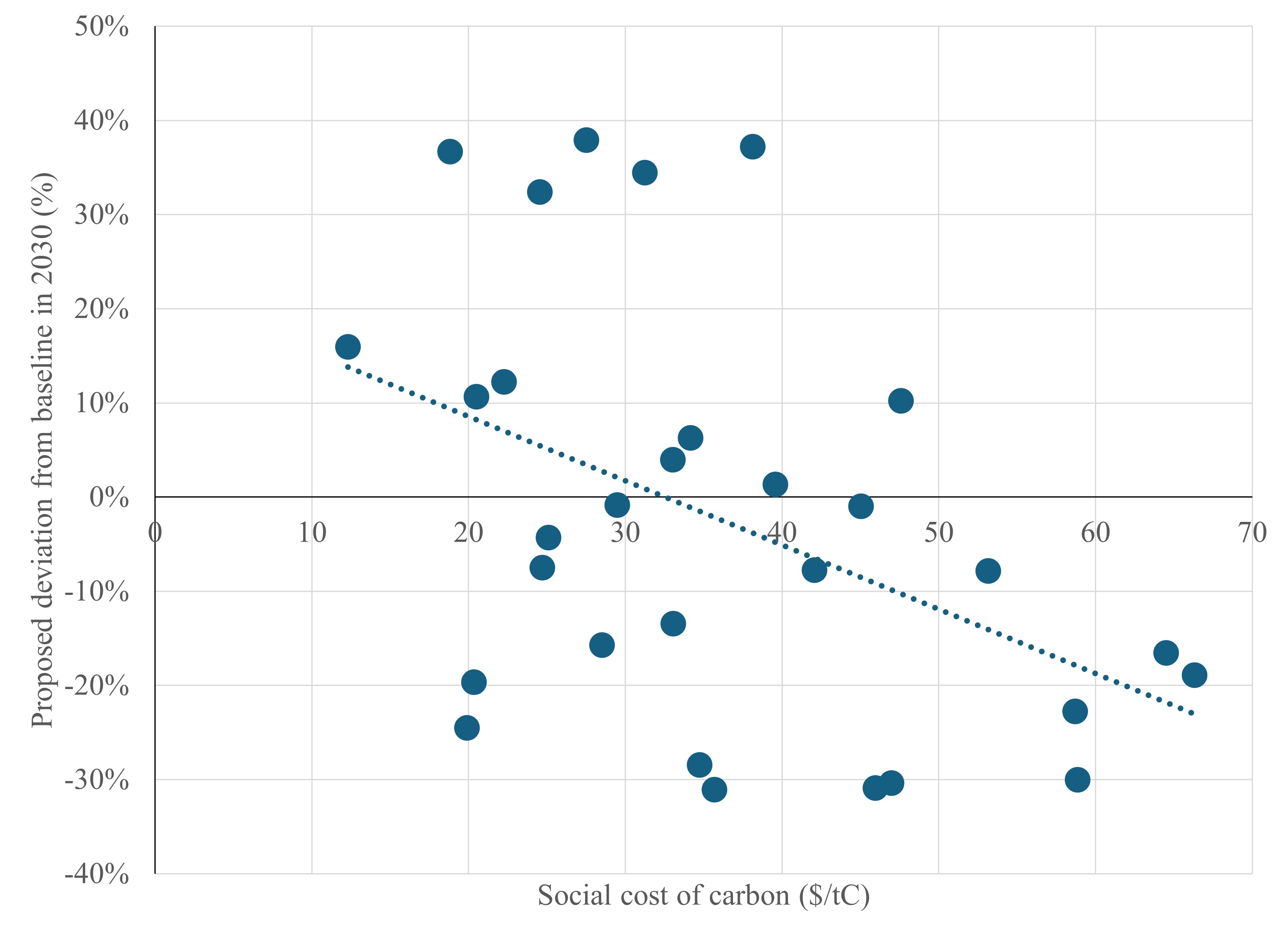}
  \includegraphics[width=0.49\textwidth]{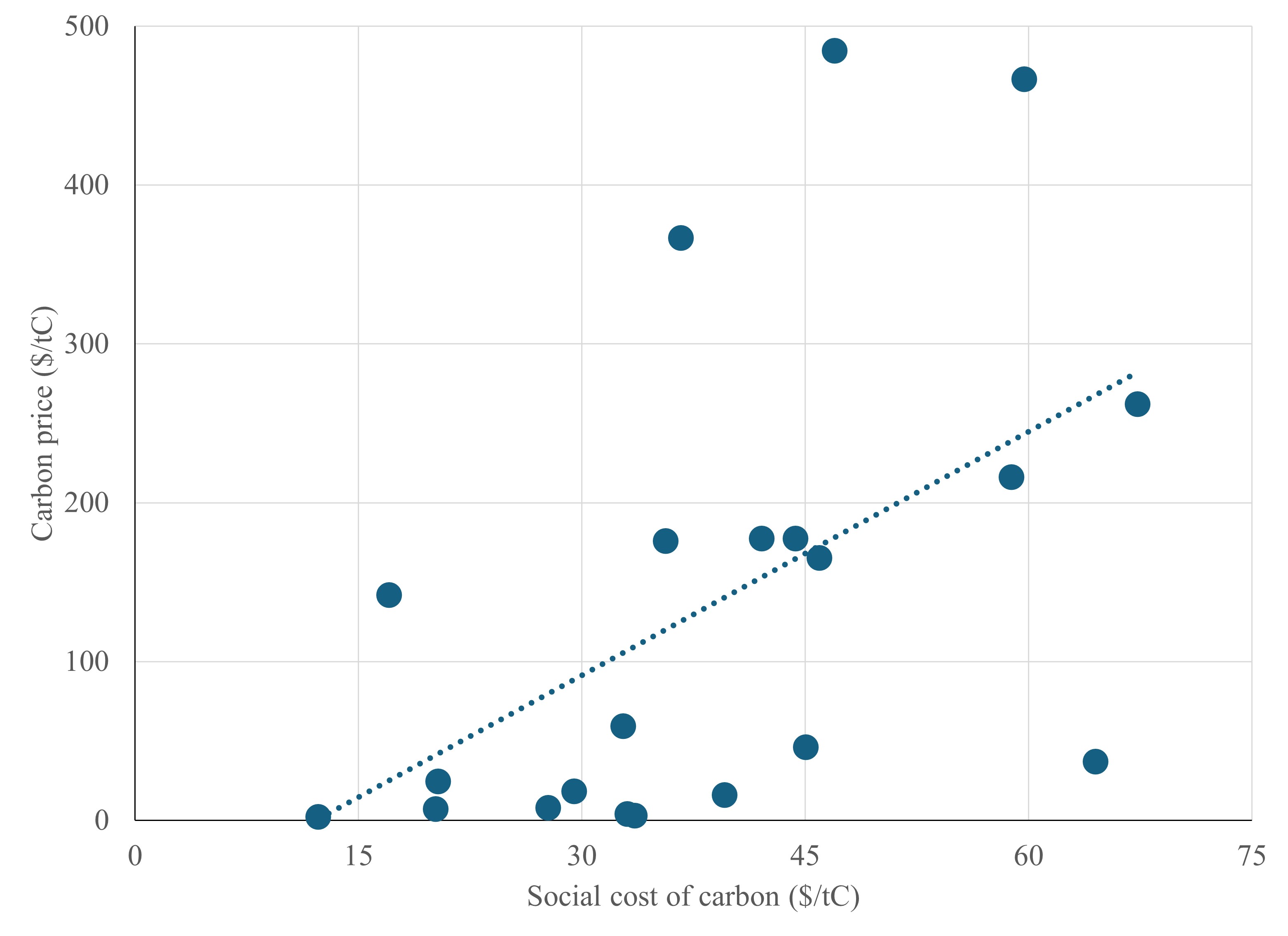}
  \includegraphics[width=0.49\textwidth]{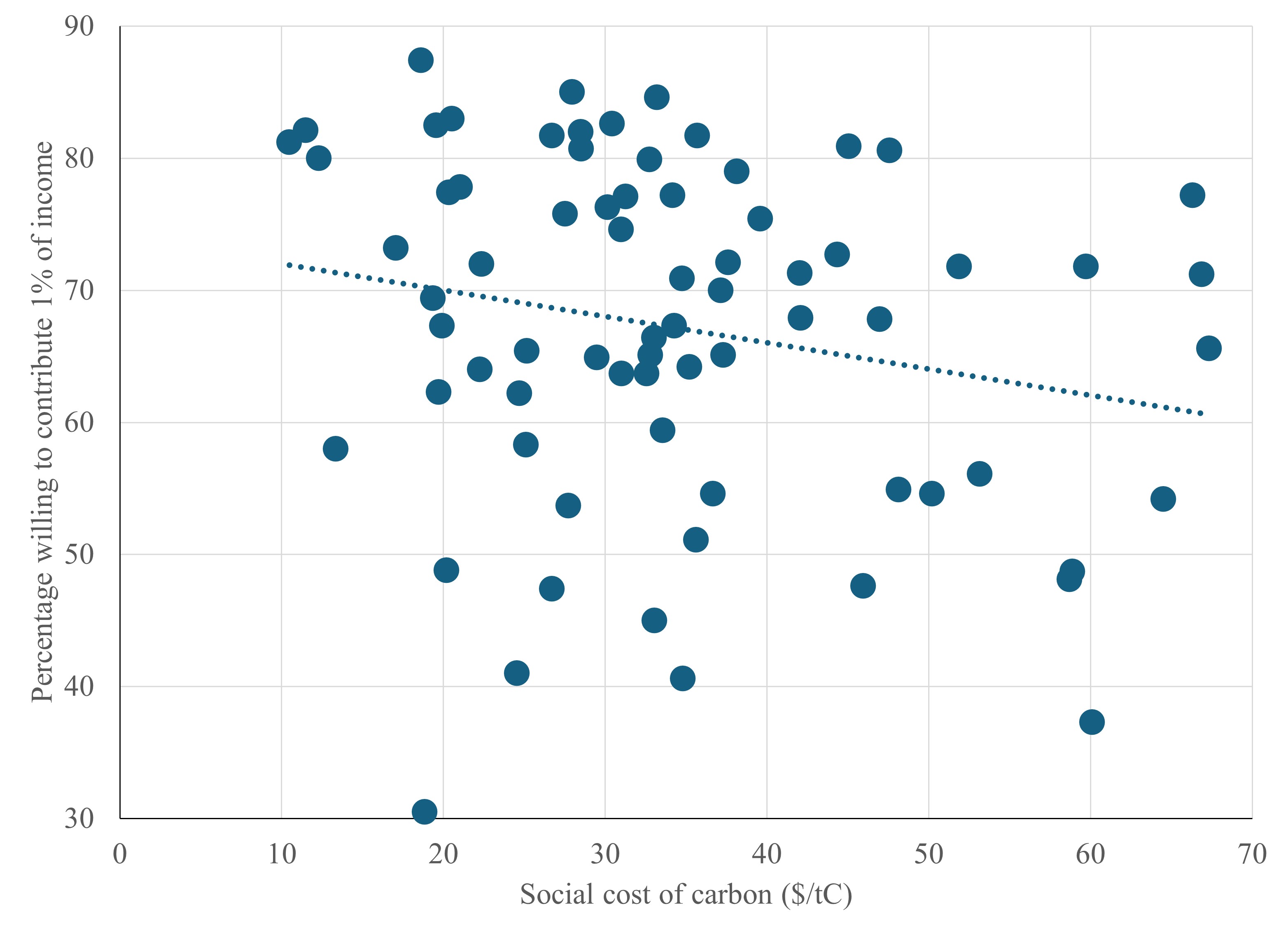}
  \includegraphics[width=0.49\textwidth]{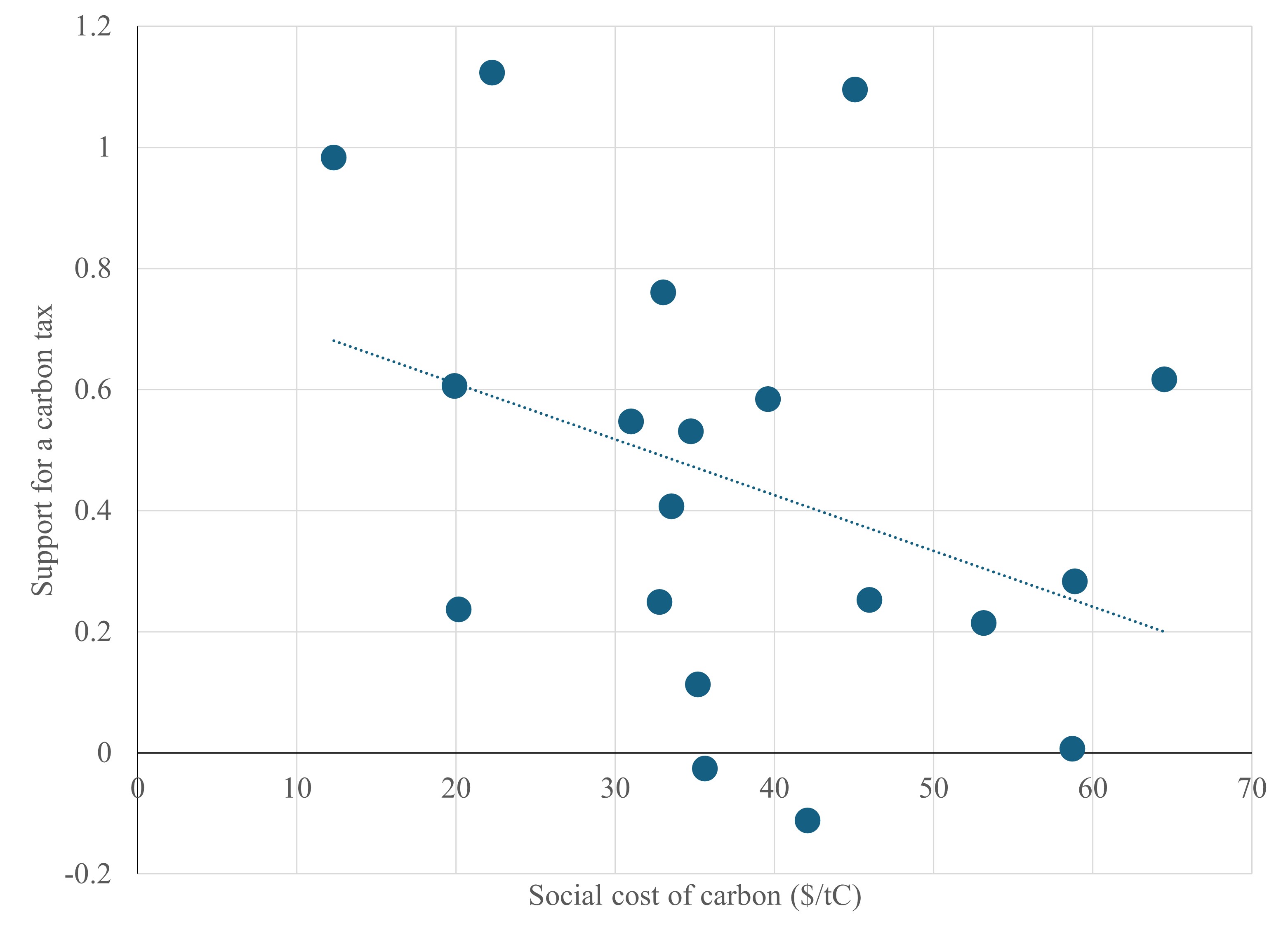}  
  \end{center}
  {\footnotesize The top left panel shows the average social cost of carbon per country plotted against the emission reduction proposed for 2030 in the Nationally Determined Contribution under the Paris Agreement of the United Nations Framework Convention on Climate Change as a percentage deviation from emissions if current policies are maintained until the end of the decade. The top right panel plots the social cost of carbon against the 2024 carbon price. The bottom left panel plots the social cost of carbon against the fraction of people willing to contribute at least one percent of their income to climate policy. The bottom right panel plots the social cost of carbon against the average support (on a scale of -2 to +2) for a carbon tax.}
\end{figure}

Cost-benefit analysis of climate policy is the remit of learned journals and economic textbooks. No country strictly aligns its climate policy to the social cost of carbon. Nonetheless, the numbers in Figure \ref{fig:premium} correlate with climate policy, as shown in Figure \ref{fig:validation}. The left panel plots the social cost of carbon with national preferences against the relative difference between the projected emissions in 2030 under current policies and the emissions as foreseen in the Nationally Determined Contributions under the Paris Agreement. The middle panel plots the social cost of carbon against the 2024 carbon price. In both cases, there is a clear albeit far from perfect correlation. The right panel plots the social cost of carbon against the fraction of people willing to contribute 1\% of income to greenhouse gas emission reduction \citep{Andre2024}; the correlation is weaker and the wrong sign. The same holds for support for a carbon tax as reported by \citet{dechez2025}. As the other two panels show the expected sign, this may be because willingness to pay tax measures capture frustration with government policy.

\section{Discussion and conclusions}
The presence of a minority who favour a low discount rate for public policy implies considerably higher investments in the long term, such as in greenhouse gas emissions reduction. While this was known qualitatively, we are the first to quantify the Weitzman premium on the social cost of carbon using a large sample of laypeople. We generalise Weitzman's gamma discounting to heterogeneity about the two parameters of the Ramsey Rule, introduce zero-inflation, and show that the analytical approximation does not capture the essence of the data. In the preferred specification, the recommended carbon price increases by a factor 5.6. In alternative specifications, the increase in the social cost of carbon ranges from 2.7 to 195. In \citet{Dong2024}, we showed that the social cost of carbon falls if we take the time preferences of people in the Global South into account. We here reverse that result. Minority views against usury matter, big time.

There are several caveats. The analytical results hinge on the gamma distribution. We explored other distributions. As far as we can see, Weitzman's choice was judicious. We merged two datasets, the Drupp data that has the right parameters (pure time preference, elasticity of intertemporal substitution) but is unrepresentative and the Falk data that is representative but has the wrong parameters (discount rate, risk aversion). We assume that short-term private preferences correlate with long-term public preferences, but we cannot test this assumption. We use a simple integrated assessment model. Particularly, we assume that the economic growth scenario used is consistent with the recalibrated time preferences of the respondents in Falk's survey. We abstract from emission reduction, from uncertainty, and from any heterogeneity except for time preferences. While much research remains to be done, this is unlikely to change the qualitative result that preference heterogeneity calls for more ambitious climate policy.

\section*{Acknowledgements}
We thank Andrea Bastianin, Gary Koop, Robert Mendelsohn, Jonathan Norris, Ivan Petrella, and Massimo Tavoni for constructive comments.

\section*{Author contributions}
JD: Conceptualization, editing; RT: Conceptualization, data, coding, writing, editing; FW: Conceptualization, editing.

\section*{Competing interest statement}
We have no financial or non-financial competing interests or other interests that might be perceived to influence the interpretation of the paper, nor does a close relative or partner.

\section*{Data availability statement}
The raw data are on the \href{https://gps.iza.org/downloads}{Global Preference Survey} site. The processed data, graphs and tables are on \href{https://www.dropbox.com/scl/fi/64vfiovggclgb4tif5l21/FalkInd.xlsx?rlkey=xb9bujwm08hxjc2s4czo0row5&st=48118y20&dl=0}{DropBox}.

\section*{Code availability statement}
The Matlab code to estimate the social cost of carbon is on \href{https://github.com/rtol/weitzmanpremium}{GitHub}.

\begin{table}[]
    \begin{center}  
    \caption{The Weitzman Premium on the social cost of carbon for alternative specifications for uncertainty about the discount rate}
    \label{tab:analytics}
    \begin{tabular}{r l c r r}
         & Data & Equation & SCC & Premium  \\ \hline
         \multicolumn{5}{c}{Samuelson discount rates} \\ \hline
         1 & Weitzman & $r = 0.0394$ & 27.3 & - \\
         2 & Weitzman & $r(t) = \frac{1.78}{45.1+t}$ & 46.5 & 70\% \\
         3 & Weitzman & empirical & 44.2 & 62\% \\
         4 & Falk & $r = 0.0708$ & 10.4 & - \\
         5 & Falk & $r(t) = \frac{2.39}{33.8+t}$ & 22.6 & 117\% \\
         6 & Falk & $r(t) = (1-0.05)\frac{2.94}{39.3+t}$ & 20.9 & 101\% \\ \hline
         \multicolumn{5}{c}{Ramsey discount rates} \\ \hline
         7 & Falk & $r(t) = 0.034+2.2g$ & 5.4 & - \\ 
         8 & Falk & $r(t) = \frac{2.66}{79.0+t} + \frac{1.08g}{0.49+tg} $ & 13.2 & 143\% \\
         9 & Falk & $r(t) = \frac{0.23 \cdot 5.00}{144.7+t} + \frac{0.10 \cdot 2.09 g}{0.79+tg}+ 0.62 \left ( \frac{6.90(1+33g)}{165+t(1+33g)} - (1.88 g)^2 t \right ) $ & $\infty$ & $\infty$ \\ 
         10 & Falk & empirical & 30.5 & 464\%      \\ \hline
    \end{tabular}
    \end{center}
    {\footnotesize Column 1 specifies the data source, either Weitzman's 2160 respondents or Falk's. Column 2 has the discount rate. The first six rows follow Samuelson, assuming that the discount rate is independent of the growth rate of the economy. The last four rows follow the Ramsey rule. Rows 1 and 4 assume a constant discount rate. Rows 2 and 5 assume Gamma discounting. Row 6 assumes zero-inflated Gamma discounting. Row 7-10 use the Ramsey rule, with constant parameters (row 7), with Gamma discounting for two independent parameters (row 8), with Gamma discounting for correlated parameters (row 9). Rows 6 and 10 use the empirical distribution. Column 3 shows the social cost of carbon for 2020, in 2015 US dollars per metric tonne of carbon. Column 4 displays the Weitzman, or taste-heterogeneity, premium}
\end{table}

\begin{table}[]
  \begin{center}
  \caption{Alternative aggregates of the social cost of carbon}
  \label{tab:global}
  \begin{tabular}{l r r r c}
  & avg. pref. & avg. SCC & Weitzman & \%ile \\ \hline
  Democracy & 5.4 & 30.5 & 25.1 & 77\\
  United Nations & 6.7 & 34.0 & 27.3 & 78\\
  Plutocracy & 7.5 & 43.1 & 35.6 & 82\\
  Welfare & 10.3 & 55.4 & 45.1 & 88\\ \hline
  \end{tabular}
  \end{center}
  {\footnotesize Row 1 shows the average social cost of carbon for the 79,273 respondents, weighted to make them representative of the world population. Row 2 first computes the population-weighted average for each of the 76 countries and then computes the unweighted average for the world. Row 3 weights the country averages with their contribution to global economic output (using market exchange rates). Row 4 equity weights the country averages, using the ratio of global average to national average per capita income (using purchasing power parity exchange rates). Column 1 shows the social cost of carbon for average preferences. Column 2 shows the average social cost of carbon for individual preference. Column 3 shows the Weitzman premium, the difference between columns 2 and 1. Column 4 shows the percentile that column 2 takes on its distribution (shown in Figure \ref{fig:histscc}).}
\end{table}

\bibliography{master}

\newpage \appendix
\setcounter{page}{1}
\renewcommand{\thepage}{A\arabic{page}}
\setcounter{table}{0}
\renewcommand{\thetable}{A\arabic{table}}
\setcounter{figure}{0}
\renewcommand{\thefigure}{A\arabic{figure}}
\setcounter{equation}{0}
\renewcommand{\theequation}{S\arabic{equation}}

\section{Additional results}

\subsection{Gamma discounting: A recap}
\citet{Weitzman2001} proposed Gamma discounting, so-called because exponential discounting combined with the assumption that the uncertainty about the discount rate follows the Gamma distribution leads to a closed-form solution for a declining discount rate. That is, if
\begin{equation}
  D(t) = e^{-rt}; f(r) = \frac{\beta^\alpha}{\Gamma (\alpha)} r^{\alpha - 1} e^{-\beta r}
\end{equation}
where $D(t)$ is the discount factor at time $t$, $r$ is the discount rate, and $f(r)$ is the probability density function of the Gamma distribution with shape parameter $\alpha$ and rate parameter $\beta$, then the certainty-equivalent discount rate $r^*$ is given by:
\begin{equation}
\label{eq:gamma}
  r^*(t) = \frac{\alpha}{\beta + t}
\end{equation}
For $t=0$, $r^* = \frac{\alpha}{\beta} = \mathbb{E} r$. For $t \rightarrow \infty$, $r^* \downarrow 0$. That is, the certainty-equivalent discount rate declines from the expectation of the discount rate to zero.

The derivation of Gamma discounting relies on the similarity between the expected discount factor, $\mathbb{E} e^{-rt}$, and the moment generating function, $\mathbb{M}(t) = \mathbb{E} e^{tr}$ (Weitzman, personal communication). The moment generating function of the Gamma distribution is
\begin{equation}
    \mathbb{E}e^{tr} = \left ( 1-\frac{t}{\beta} \right )^{-\alpha} = \left ( \frac{\beta}{\beta-t} \right )^\alpha
\end{equation}
Replace $t$ with $-t$ to find the expected discount factor. The expected discount rate is the growth rate of the discount factor
\begin{equation}
\label{eq:gamma}
    \mathbb{E}r = - \frac{\dot{D}}{D} = -\frac{\alpha \left (\frac{\beta}{\beta+t} \right )^{\alpha-1} \frac{-\beta}{(\beta+t)^2}}{\left(\frac{\beta}{\beta + t} \right )^\alpha} = \frac{\alpha}{\beta + t}
\end{equation}

\subsection{Zero-inflation}
For alternative assumptions about the distribution of the discount rate, the certainty-equivalent discount rate declines too, but not according to a simple equation or even a closed-form function.

\begin{figure}
  \begin{centering}
  \caption{Histogram and density of individual discount rates}
  \label{fig:histdr}
  \includegraphics[width=\textwidth]{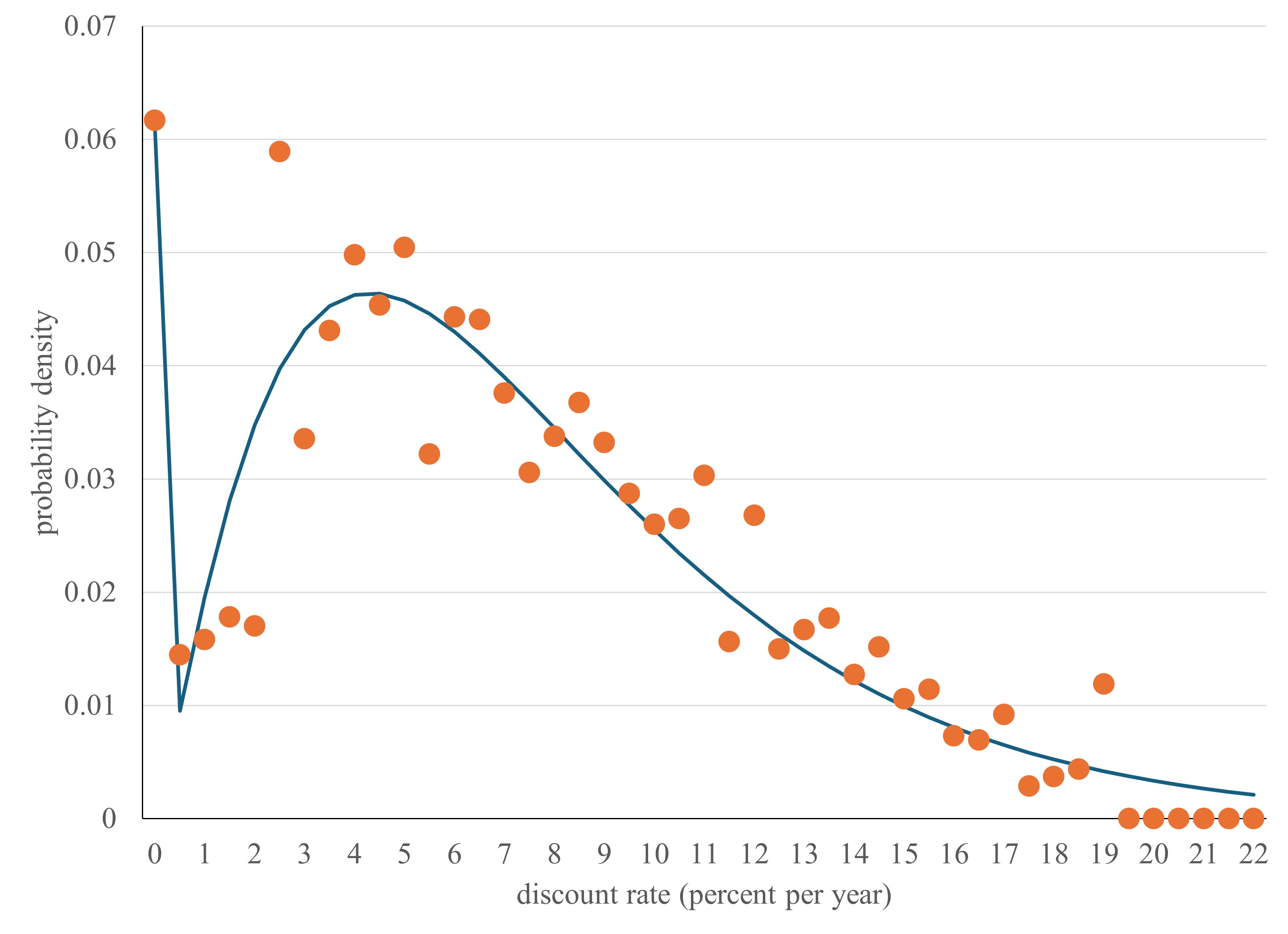}
  \end{centering}
  {\footnotesize The orange dots show the histogram of individual Ramsey discount rates, assuming a consumption growth rate of 2\% per year, of the 79,272 respondents. Observations are weighted to be representative of the world population. The blue line shows a zero-inflated gamma distribution, fitted to the observations using mean integrated squared error.}
\end{figure}

Figure \ref{fig:histdr} plots the histogram of the discount rate for the 79,272 respondents. The discount rate is assumed to be equal to their calibrated pure rate of time preference plus their calibrated elasticity of the marginal utility of consumption times 1.7, the global average growth rate of per capita consumption between 2014 and 2023.

Figure \ref{fig:histdr} also shows a fitted Gamma distribution, where parameters $\alpha = 2.24$ and $\beta = 0.29$ were chosen to minimize the mean integrated squared error. Or rather, because some 6\% of respondents prefer a \emph{zero} discount rate, the graph shows a \emph{zero-inflated} Gamma distribution, with a $\delta = 0.06$ probability mass at zero. It is a reasonable fit, but not a perfect one.

The zero-inflated gamma discount rate is
\begin{equation}
  r^\dag(t) = \delta \times 0 + (1-\delta)\frac{\alpha}{\beta + t} = (1-\delta)\frac{\mu}{1 + t \sfrac{\sigma^2}{\mu}}
\end{equation}
where $\mu=0.08$ denotes the expected value and $\sigma^2=0.64$ the variance. With the parameters above, the initial annual discount rate is 7.28\%. It falls to 0.02\% per year in 100 years.

\begin{figure}
  \begin{centering}
  \caption{Alternative discount rates}
  \label{fig:altdr}
  \includegraphics[width=\textwidth]{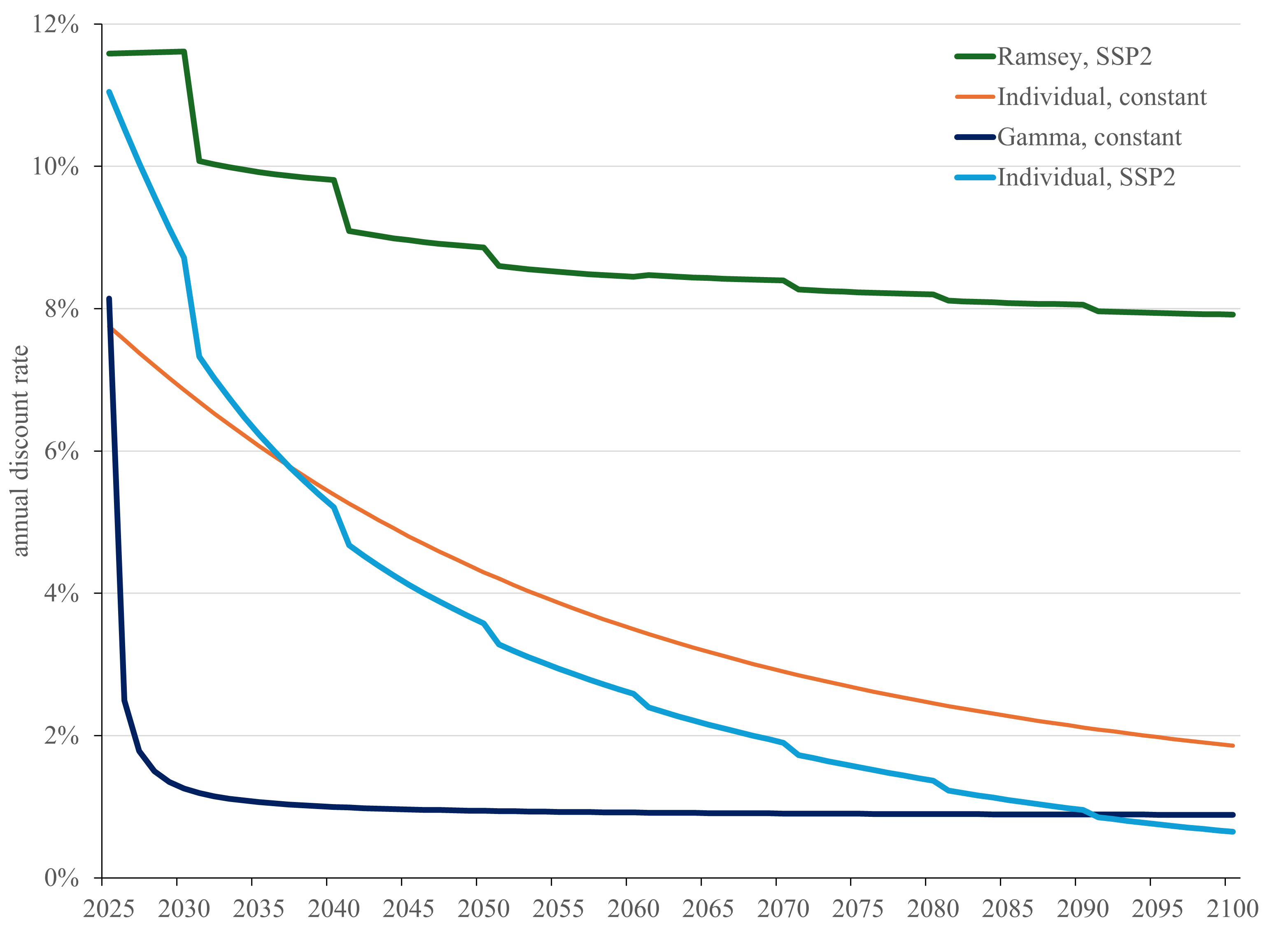}
  \end{centering}
  {\footnotesize From top to bottom (in 2100), the Ramsey discount rate for population and income growth according to the SSP2 scenario and global average PRTP and EMUC; the certainty-equivalent discount rate for the 79,272 respondents and constant population and income growth; the zero-inflated gamma discount rate with constant population and income growth; and the certainty-equivalent discount rate with the SSP2 scenario.}
\end{figure}

At a \emph{constant} discount rate of 7.28\% plus the average population growth rate of 0.85\%, the social cost of carbon is \$9.8/tC. This is higher than the social cost of carbon in Table \ref{tab:global}, because observed population growth has been slower than projected in the first decade of the SSP2 scenario. Figure \ref{fig:altdr} confirms that a discount rate calibrated to the economic growth of the recent past is \emph{lower} than one calibrated to the near future according to SSP2.

With a \emph{declining} discount rate\textemdash Equation (\ref{eq:gamma})\textemdash the social cost of carbon increases to \$145.0/tC. Weitzman's analytical approximation overstates his premium by a considerable margin. There is a positive correlation between time and risk, which is considered when estimating the individual social costs of carbon, but not in the approximation with gamma discounting. This puts too little weight on very impatient respondents. However, this effect is dominated by the rapid decline in the discount rate. 

Figure \ref{fig:altdr} illustrates this. The gamma discount rate falls almost instantly to its lower limit, i.e., the population growth rate. This is because $\sfrac{\sigma^2}{\mu} = 8.0$ so that Equation (\ref{eq:gamma}) is dominated by $t$.\footnote{Note that $\sfrac{\sigma^2}{\mu}$ is not unitless. It would be 800 if measured in percentage per year rather than fraction per year.} In other words, Weitzman's analytical approximation with the gamma distribution leads to a discount rate that declines too rapidly in case of overdispersion.

\subsection{Normality}

For the Normal or Gaussian distribution
\begin{equation}
    \mathbb{E}r = - \frac{\dot{D}}{D} = -\frac{e^{-\mu t + 0.5 \sigma^2 t^2} (-\mu + \sigma^2 t)}{e^{-\mu t + 0.5 \sigma^2 t^2}} =  \mu - \sigma^2 t
\end{equation}
This implies \emph{negative} discount rates in the long run.

\subsection{Risk aversion}
In the Ramsey rule, the curvature of the utility function is multiplied by the growth rate of consumption. The moment generating function $\mathbb{M}(t)$ of $ar+b$ is
\begin{equation}
    \mathbb{M}_{ar+b}(t) = e^{bt} \mathbb{M}_r(at)
\end{equation}
Therefore, if we assume that the inverse of the intertemporal elasticity of substitution is Gamma distributed, then
\begin{equation}
    \mathbb{E}e^{- t \eta g} = \left ( \frac{\beta}{\beta+tg} \right )^\alpha
\end{equation}
The expected discount rate is thus
\begin{equation}
    \mathbb{E}r = - \frac{\dot{D}}{D} = -\frac{\alpha \left (\frac{\beta}{\beta+tg} \right )^{\alpha-1} \frac{-\beta g}{(\beta+tg)^2}}{\left(\frac{\beta}{\beta + tg} \right )^\alpha} = \frac{\alpha g}{\beta + tg}
\end{equation}

\subsection{Independent time and risk}
The Ramsey rule has that the discount rate equals the pure rate of time preference plus the growth rate of consumption times the curvature of the utility function. The moment generating function of the sum of two \emph{independent} random variables is the product of the moment generating functions. The expected discount factor is thus
\begin{equation}
    \mathbb{E}e^{-(\rho + \eta g)t} = \left ( \frac{\beta_\rho}{\beta_\rho+t} \right )^{\alpha_\rho} \left ( \frac{\beta_\eta}{\beta\eta+gt} \right )^{\alpha_\eta}
\end{equation}
The expected discount rate is then
\begin{equation}
\label{eq:indgamma}
    \mathbb{E}r = \mathbb{E} (\rho + \eta g) = \frac{\alpha_\rho }{\beta_\rho + t} + \frac{\alpha_\eta g}{\beta_\eta + tg}
\end{equation}
This behaves much like the original Gamma discounting. At $t=0$, the expected discount rate is the expected pure rate of time preference plus the expected rate of relative risk aversion times the growth rate of consumption. As $t$ increases, the expected discount rate falls. However, its two components fall at different rates. Together, they fall more slowly than Equation (\ref{eq:gamma}).

To see that, rewrite Equation (\ref{eq:gamma}) as
\begin{equation*}
    r^{1D}:= \frac{\alpha_r}{\beta_r + t} = \frac{\mu_r}{1+t\frac{\sigma_r^2}{\mu_r}} = \frac{\mu_\rho + \mu_\eta g}{1+t\frac{\sigma_\rho^2 + g^2 \sigma_\eta^2}{\mu_\rho + \mu_\eta g}}
\end{equation*}
and Equation (\ref{eq:indgamma}) as
\begin{equation*}
r^{2D}:= \frac{\alpha_\rho }{\beta_\rho + t} + \frac{\alpha_\eta g}{\beta_\eta + tg} = \frac{\mu_\rho}{1 + t \frac{\sigma_\rho^2}{\mu_\rho}} + \frac{\mu_\eta g}{1 + tg \frac{\sigma_\eta^2}{\mu_\eta}} = \frac{\mu_\rho + \mu_\eta g + gt \left ( \mu_\rho \frac{\sigma_\eta^2}{\mu_\eta} + \mu_\eta \frac{\sigma_\rho^2}{\mu_\rho} \right )}{1 + t \frac{\sigma_\rho^2}{\mu_\rho} + tg \frac{\sigma_\eta^2}{\mu_\eta} + t^2 g \frac{\sigma_\rho^2 \sigma_\eta^2}{\mu_\rho \mu_\eta}}
\end{equation*}
Cross-multiply, expand and simplify to find
\begin{equation*}
r^{2D}-r^{1D} =
\frac{tg(\mu_\rho+\mu_\eta g)}
{\left(1+t\frac{\sigma_\rho^2}{\mu_\rho}\right)
\left(1+tg\frac{\sigma_\eta^2}{\mu_\eta}\right)
\bigl(\mu_\rho+\mu_\eta g+t(\sigma_\rho^2+g^2\sigma_\eta^2)\bigr)}
\left(
\mu_\rho\frac{\sigma_\eta^2}{\mu_\eta}
-\mu_\eta\frac{\sigma_\rho^2}{\mu_\rho}
\right)^2.
\end{equation*}
All parameters are positive so $r^{1D}<r^{2D}$\textemdash unless $\mu_\rho^2\sigma_\eta^2=\mu_\eta^2\sigma_\rho^2$ or $t=0$, in which case $r^{1D}=r^{2D}$. In words, Weitzman's one-dimensional Gamma discount rate is no higher than its two-dimensional generalization. The one-dimensional gamma discount factor falls at least as fast as its two-dimensional counterpart.

\subsection{Correlated time and risk}
Time and risk preferences are typically correlated, but bivariate Gamma distributions are not easy to handle \citep{Nadarajah2006}.

Auxiliary regression removes correlation. Regress $\eta$ on $\rho$ to find $\eta = \kappa \rho + \varepsilon$. Then, $\rho$ and $\varepsilon$ are uncorrelated and $r = \rho + \eta g = \rho (1+\kappa g) + \varepsilon g$.

As above with two independent Gamma distributions, the expected discount rate of the sum of a Gamma and Normal distribution is the sum of the expected discount rates. Therefore,  
\begin{equation}
    \mathbb{E}r = \frac{\alpha_\rho (1+\kappa g)}{\beta_\rho + t(1 + \kappa g)} - \sigma_\varepsilon^2 g^2 t
\end{equation}

\subsection{Conclusion}
We thus have four regimes:
\begin{equation}
r(t) =
\begin{cases}
      0 & \text{if } \rho=0, \eta=0 \\
     \frac{\alpha_\rho}{\beta_\rho + t} & \text{if } \rho>0, \eta=0 \\
     \frac{\alpha_\eta g}{\beta_\eta + gt} & \text{if } \rho=0, \eta>0 \\
     \frac{\alpha_\rho (1+\kappa g)}{\beta_\rho + t(1 + \kappa g)} - \sigma_\varepsilon^2 g^2 t & \text{if } \rho>0, \eta>0
\end{cases}
\end{equation}

\begin{figure}
    \centering
    \caption{The cumulative distribution function of the social cost of carbon for the time preference parameters in the Drupp survey and the parameters calibrated to the time and risk indices for the respondents to the Falk survey in North America and Western Europe.}
    \label{fig:calib}
    \includegraphics[width=\textwidth]{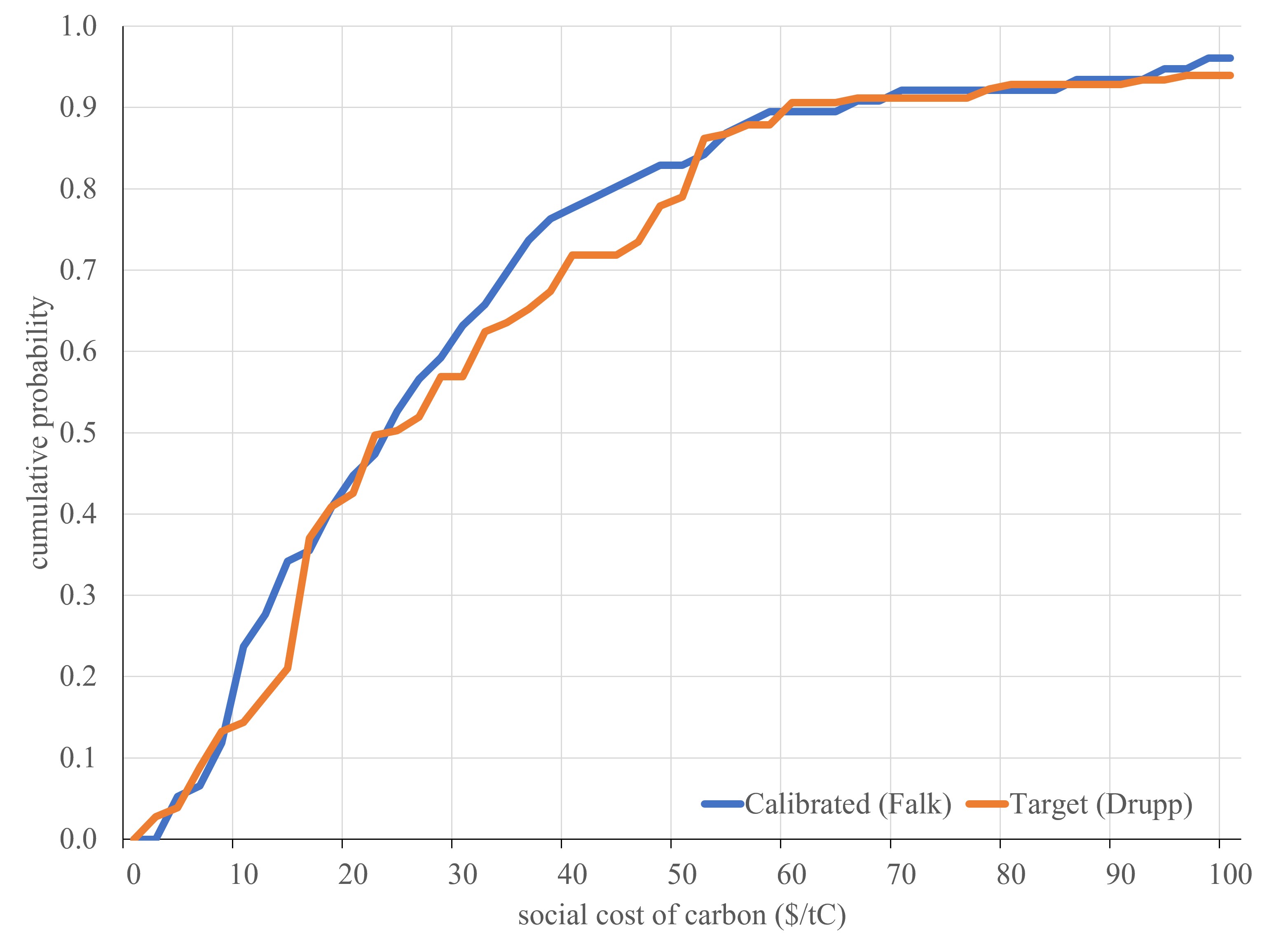}
    \caption*{After \citet{Dong2024}.}
\end{figure}

\begin{figure}
    \centering
    \caption{The histogram of the consumption discount rate for independent and correlated time and risk preferences}
    \label{fig:correl}
    \includegraphics[width=\textwidth]{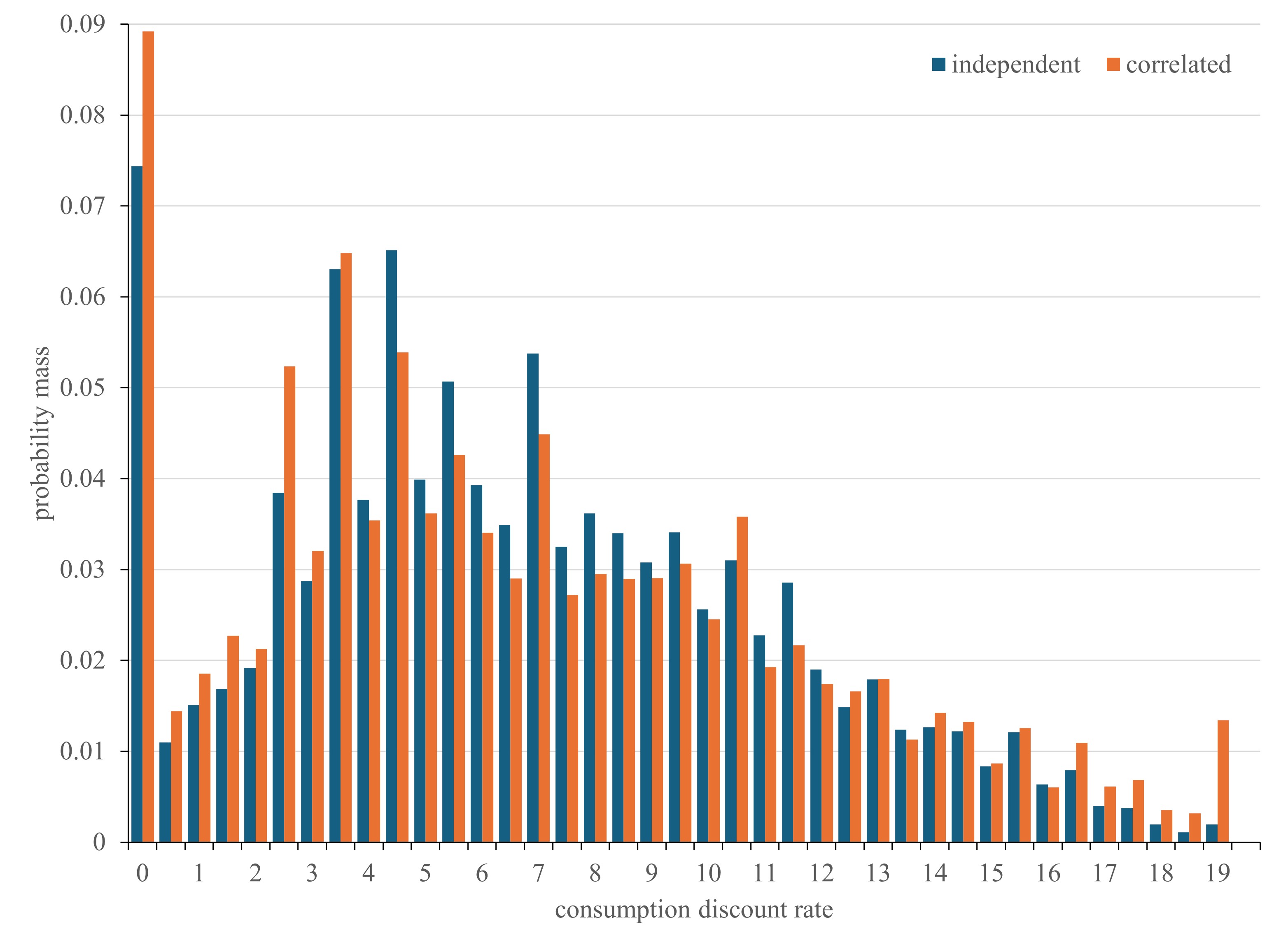}
\end{figure}

\begin{figure}
  \begin{centering}
  \caption{An illustration of a declining discount rate}
  \label{fig:gamma}
  \includegraphics[width=\textwidth]{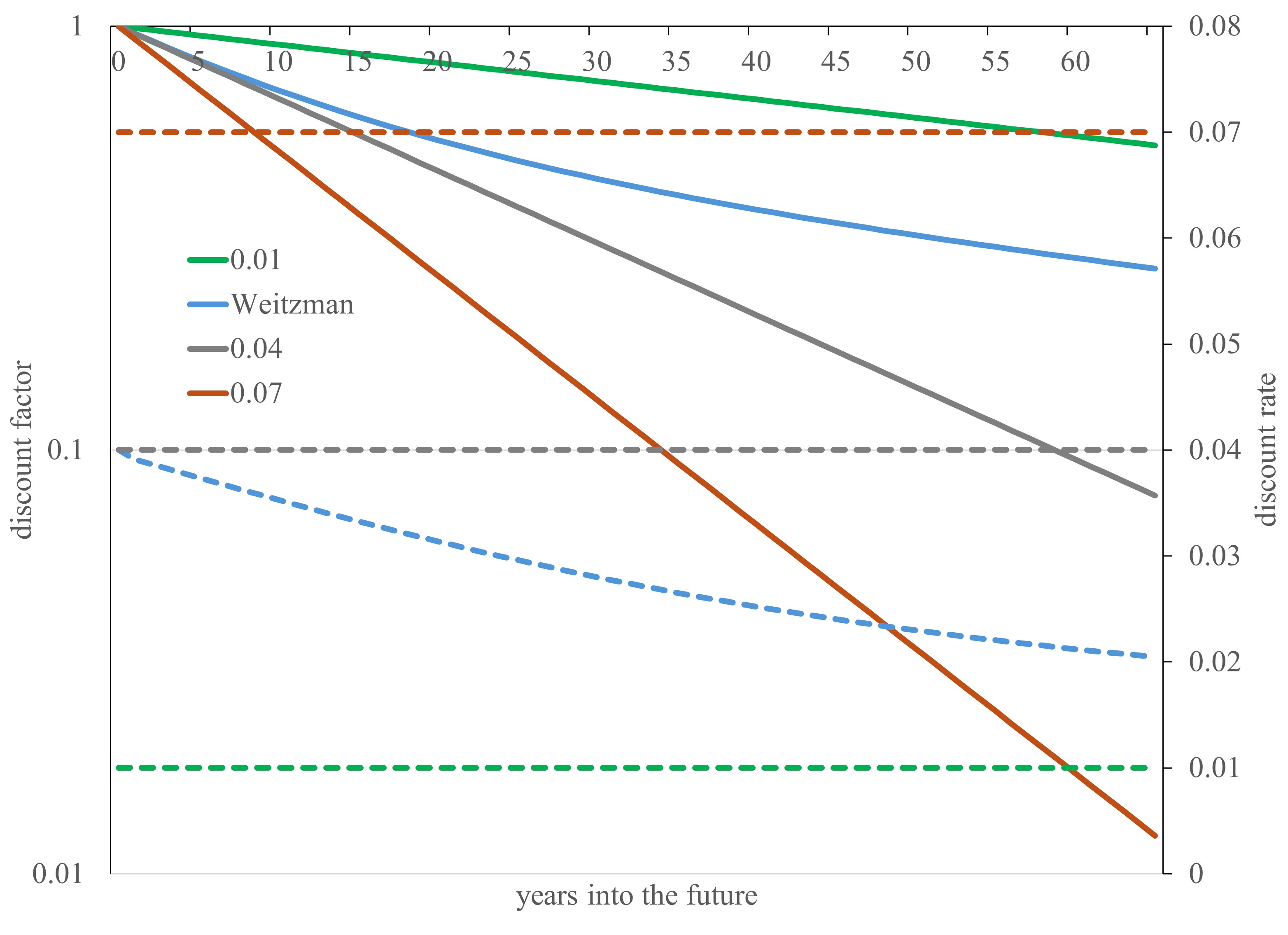}
  \end{centering}
  {\footnotesize Four alternative discount rates $r$ are displayed on the right axis in dashed lines: 1\% per year, 7\%, the average of 1\% and 7\% (4\%), and the corresponding declining discount rate (Weitzman), which starts at 4\% and falls to 2\% after 65 years. It falls to 1.1\% after 1000 years and to 1\% at eternity (not shown). The corresponding discount factors, $(1+r)^{-t}$, are shown on the left axis in solid lines. The discount factor is the weight placed on future impacts. A constant discount rate implies that the discount factor appears as a straight line on a logarithmic scale as used here. The declining discount factor is at first almost parallel to the discount factor for the average discount rate $r=0.04$ but later is almost parallel to the discount factor for the minimum discount rate $r=0.01$.}
\end{figure}

\end{document}